\documentclass[twocolumn,preprintnumbers,amsmath,amssymb,superscriptaddress,prd]{revtex4-2}

\usepackage{graphicx}

\newcommand{\aap}{Astron. Astrophys.}

\newcommand{\apjl}{Astrophys. J. Lett.}

\def\aap{Astron. Astrophys.}
\def\prl{Phys. Rev. Lett.}
\def\prd{Phys. Rev. D}
\def\apjl{Astrophys. J. Lett.}

\usepackage{dcolumn}
\usepackage{bm}
\usepackage{hyperref}
\usepackage{natbib}
\usepackage{color}

\begin{document}

\title{On the Apparent Correlation between X-ray and Neutrino Luminosities of Active Galactic Nuclei}

\author{Luo, Jian-Jun}
\affiliation{Guangxi Key Laboratory for Relativistic Astrophysics, School of Physical Science and Technology, Guangxi University, Nanning 530004, People's Republic of China}
\affiliation{GXU-NAOC Center for Astrophysics and Space Sciences, Nanning 530004, People's Republic of China}
\author{Lu, Ming-Xuan}
\email[]{lumx@st.gxu.edu.cn}
\affiliation{Guangxi Key Laboratory for Relativistic Astrophysics, School of Physical Science and Technology, Guangxi University, Nanning 530004, People's Republic of China}
\affiliation{GXU-NAOC Center for Astrophysics and Space Sciences, Nanning 530004, People's Republic of China}
\author{Liang, Yun-Feng}
\affiliation{Guangxi Key Laboratory for Relativistic Astrophysics, School of Physical Science and Technology, Guangxi University, Nanning 530004, People's Republic of China}
\affiliation{GXU-NAOC Center for Astrophysics and Space Sciences, Nanning 530004, People's Republic of China}

\date{\today}

\begin{abstract}
	Recent studies have reported a linear correlation between the hard X-ray and high-energy neutrino luminosities of active galactic nuclei (AGN), suggesting a possible physical connection between these two messengers. In this work, we challenge this interpretation by demonstrating that the observed correlation may arise purely from selection effects. We analyze 10 years of IceCube public data for a sample of Seyfert galaxies and blazars from the \textit{Swift} BAT catalog. While our data reproduces the apparent $L_\nu$--$L_X$ correlation for sources with mild (but not significant) neutrino evidence, we show through Monte Carlo simulations that the same correlation appears even when analyzing random sky positions with no astrophysical sources. The key issue is that TS-based source selection effectively restricts the neutrino flux to a narrow range (a factor of several), while the luminosity distance of the sample spans $\sim4$ orders of magnitude. This causes the luminosity $L = 4\pi D_L^2 F$ to be dominated by the distance term rather than intrinsic flux variations, creating an artificial correlation.
While a robust flux correlation ($F_\nu$--$F_X$) for high-significance sources may indicate a genuine physical link, our results demonstrate that a luminosity-luminosity correlation alone is insufficient to establish a physical relationship between neutrino and X-ray emission in AGN.
\end{abstract}

\maketitle

\section{Introduction}
\label{sec:intro}

The IceCube Neutrino Observatory has detected a diffuse flux of high-energy astrophysical neutrinos
in the TeV--PeV energy range \citep{Aartsen2013}, opening a new window for multimessenger astronomy.
Despite more than a decade of observations, the sources contributing to this diffuse flux remain
largely unidentified. Active galactic nuclei (AGN), with their powerful central engines and dense
radiation fields, have long been considered promising candidates for high-energy neutrino production
\citep{Halzen1997, Mannheim1995}.

The first compelling evidence for neutrino emission from an individual AGN came from the blazar TXS 0506+056, identified through a time-dependent correlation with a high-energy neutrino alert in 2017 \cite{Aartsen2018_TXS,IceCube:2018cha}. The first evidence for steady neutrino emission from an AGN came from the nearby Seyfert galaxy NGC 1068, which remains the most significant neutrino point source detected by IceCube to date \cite{Aartsen2020_PRL,Abbasi2022}.
\citet{Aartsen2020_PRL} reported an excess of $2.9\sigma$ post-trial significance from the direction of
NGC 1068 using 10 years of IceCube data, which was later confirmed with updated analyses reaching
$\sim5\sigma$ local significance ($\sim4.2\sigma$ global significance) \citep{Abbasi2022}. 
The observed neutrino
flux from NGC 1068 exceeds its $\gamma$-ray flux by over one order of magnitude, suggesting that neutrinos
are produced in a $\gamma$-ray obscured environment, likely the AGN corona near the supermassive
black hole \citep{Murase2020,Inoue2020,Murase2014,Eichmann:2022lxh}. This discovery motivated systematic searches for neutrino
emission from other X-ray bright AGN \cite{IceCube:2024ayt,Abbasi2024a,IceCube:2025gdd,Abbasi2026}.

Many works have extended this search to larger samples of Seyfert galaxies
and AGN. IceCube has performed extensive time-integrated searches for neutrino point sources using muon track events
\citep{Aartsen2020_PRL}. 
\citet{IceCube:2024ayt} performed a search using 12 years of all-sky IceCube muon track data (PSTracks) on 43 hard X-ray AGN from the Swift-BAT catalog, reporting no significant stacking signal but identifying an excess from NGC 4151 at the $2.9\sigma$ post-trial level. \citet{Abbasi2024a} analyzed 10 years of Northern Hemisphere through-going track data on 27 X-ray bright Seyfert galaxies, finding a $2.7\sigma$ post-trial significance in a binomial test driven by excesses from NGC 4151 and CGCG 420-015. More recently, \citet{IceCube:2025gdd} extended the Northern track analysis to 13.1 years of data and reported a $3.3\sigma$ collective excess from 11 X-ray bright AGN (excluding NGC 1068), with NGC 1068 itself reaching $4.0\sigma$ global significance. In the Southern Hemisphere, \citet{Abbasi2026} utilized the enhanced starting track event selection (ESTES) on 10 years of data (2011--2021) to report a $3.0\sigma$ stacking signal from 14 Seyfert galaxies.
Searches for neutrino emission from radio-bright AGN and gamma-ray blazars have also been conducted \citep{Zhou2021,2022PhRvD.106h3024L,IceCube:2023htm}.

Based on the IceCube observations, \citet{Kun2024} proposed a possible linear correlation between
the unabsorbed hard X-ray luminosity ($L_{\rm X}$) and the neutrino luminosity ($L_\nu$) for AGN.
Analyzing six sources (four Seyfert galaxies and two blazars), they found a Pearson correlation
coefficient of $R = 0.97$, suggesting a common neutrino production mechanism in $\gamma$-ray obscured
regions near the central black hole. This correlation was later extended to eight AGN by
Ref.~\cite{Kun2025}, where the Seyfert galaxy NGC 5610 is identified as a hidden contributor to the neutrino
hotspot near the blazar PKS 1424+240, strengthening the correlation ($R_P = 0.919$). These works
proposed that the hard X-ray flux, produced through cascade reprocessing of $\gamma$-rays from
pion decay, should be comparable to the neutrino flux, providing a testable prediction for
identifying neutrino source candidates.

However, the origin of this apparent $L_\nu$--$L_{\rm X}$ correlation remains to be
determined. The current sample of neutrino source candidates is limited to sources with modest
statistical significance (typically $\sim3$--$5\,\sigma$ local significance), and the correlation has only been tested
on a small number of sources ($< 10$). Furthermore, the selection of neutrino candidates
itself imposes constraints on the observable flux range, which may introduce biases in the
luminosity-luminosity plane when combined with the large spread in source distances.

In this work, we examine the claimed linear correlation between X-ray and neutrino
luminosities using a large sample of 964 Seyfert galaxies and 150 blazars from the Swift-BAT
catalog. We analyze IceCube's 10-year public dataset to search for neutrino emission from these sources. While our observational data initially show
an apparent linear correlation similar to that reported by Refs.~\cite{Kun2024,Kun2025}, our further investigation and simulations demonstrate that this correlation may arise as a selection effect rather than
reflecting intrinsic physical properties of the sources.

\section{Data and Analysis}
\label{sec:data}

\subsection{Source Sample}

Our source sample is drawn from the \textit{Swift} BAT 157-month hard X-ray survey catalog\footnote{\url{https://swift.gsfc.nasa.gov/results/bs157mon/}} \citep{Lien:2025zzo}. This catalog represents the most comprehensive hard X-ray survey from Swift/BAT to date, incorporating an additional 4.5 years of data beyond the previous 105-month catalog, with uniformly reprocessed data and updated instrumental calibration. The survey reaches a sensitivity of $8.83 \times 10^{-12}$ erg s$^{-1}$ cm$^{-2}$ for 90\% of the sky and $6.44 \times 10^{-12}$ erg s$^{-1}$ cm$^{-2}$ for 10\% of the sky in the 14--195 keV band.

The full catalog contains 1,888 sources, including 514 Seyfert I galaxies, 462 Seyfert II galaxies, and 184 beamed AGN (blazars/FSRQs). From this master catalog, we select two classes of AGN that have been proposed as potential neutrino emitters and have been shown to satisfy the $L_\nu-L_X$ relation: Seyfert galaxies and blazars. We further require sources to have measured redshifts. After applying these selection criteria, our final sample contains 964 Seyfert galaxies and 150 blazars.
For each source, we obtain the X-ray flux from the BAT catalog in the 14--195 keV band.

To compare with previous studies that used the 15--55 keV band, we convert the fluxes adopting a power-law spectrum with the photon index $\Gamma$ reported in the catalog. We note that the hard X-ray emission mechanisms in our sample differ by source class, typically originating from thermal Comptonization in the coronae of Seyfert galaxies and non-thermal jet emission in blazars. However, over the narrow observed energy range, these continuum spectra can be reasonably approximated by a power-law spectrum. Though individual sources may exhibit spectral complexity such as absorption, reflection, or curvature; for a large-sample statistical analysis, the power-law approximation is sufficient.

The X-ray luminosity is then calculated as:
\begin{equation}
	L_X = 4\pi D_L^2 \cdot F_X \cdot K,
\end{equation}
where $D_L$ is the luminosity distance (computed from the redshift using standard $\Lambda$CDM cosmology with $H_0 = 67.4$ km s$^{-1}$ Mpc$^{-1}$, $\Omega_m = 0.315$, $\Omega_\Lambda = 0.685$ \cite{Planck2018}), $F_X$ is the flux in the 15--55 keV band, and $K$ is the k-correction factor accounting for the bandpass shift due to redshift. Since the sources in the sample have small redshifts, the k-correction has a negligible effect on the results. While previous works \citep{Kun2024,Kun2025} used NuSTAR observations to derive more accurate X-ray fluxes, the 15--55 keV fluxes derived from the BAT catalog are sufficient for our purpose of demonstrating the selection effect.
\subsection{IceCube Neutrino Analysis}

We analyze 10 years of public IceCube data (2008--2018) \cite{IceCube:2021xar} using the \texttt{skyllh} software package \citep{IceCube:2023ihk,skyllh_website}, an open-source Python-based maximum-likelihood analysis framework developed by the IceCube Collaboration for point-source searches. The data span the period from April 6, 2008 to July 8, 2018, encompassing multiple detector configurations: IC40, IC59, and IC79 during the construction phases with 40, 59, and 79 strings, respectively, followed by IC86-I through IC86-VII with the full 86-string array \citep{Aartsen2020_PRL,IceCube:2021xar}. The data consist primarily of through-going muon neutrino track events, which have superior angular resolution (median $\lesssim 1^\circ$) compared to cascade events and are well-suited for point-source analysis at TeV energies. For seasons IC86-II through IC86-VII, multi-variate Boosted Decision Trees (BDT) are employed to reduce atmospheric background, preserving $\sim$90\% of atmospheric neutrinos while rejecting $\sim$99.9\% of atmospheric muons in the Northern hemisphere \citep{IceCube:2021xar}. The final all-sky event rate is approximately 4 mHz, dominated by atmospheric muon neutrinos in the Northern hemisphere and high-energy atmospheric muons in the Southern hemisphere. The angular reconstruction has been updated in v3, achieving more than 10\% improvement in angular resolution for events above 10 TeV compared to previous versions \citep{IceCube:2021xar}.

For each X-ray AGN in our sample, we perform a point-source likelihood analysis at the source position reported in the catalog. The likelihood function is defined as \cite{Braun:2008bg, Braun:2009wp}:
\begin{equation}
	\mathcal{L}(n_s, \gamma) = \prod_{i=1}^{N} \left[ \frac{n_s}{N} S_i(\gamma) + \left(1 - \frac{n_s}{N}\right) B_i \right],
\end{equation}
where $n_s$ is the number of signal events, $\gamma$ is the spectral index, $S_i$ is the signal probability density function (PDF) for event $i$ (which depends on the event's direction reconstruction uncertainty and the source position), and $B_i$ is the background probability density function. The background PDF is estimated from the data itself, based on the fact that only a very small number of candidate neutrino sources have been found to date in the muon data.
For further details on the likelihood analysis, see the skyllh documentation \cite{IceCube:2023ihk,skyllh_website} and our previous works \cite{2022PhRvD.106h3024L,2023PhRvD.108d3001G,2026arXiv260221542O}.

\begin{table*}[t]
\centering
\caption{Top 10 sources ranked by the TS values from our point-source likelihood analysis of 1114 AGN in the \textit{Swift} BAT catalog using IceCube neutrino data.}
\label{tab:top10}
\begin{ruledtabular}
\begin{tabular}{llcccc}
BAT Name            & Counterpart             & TS    & $n_s$ & $L_\nu$ (erg s$^{-1}$) & $z$    \\
\hline
SWIFT J0242.6+0000  & NGC 1068                & 19.77 & 56.72 & $5.90 \times 10^{42}$  & 0.0038 \\
SWIFT J1210.5+3924  & NGC 4151                & 18.38 & 42.71 & $1.72 \times 10^{42}$  & 0.0033 \\
SWIFT J0202.4+6824B & 2MASX J02013241+6824219 & 15.41 & 44.63 & $4.90 \times 10^{43}$  & 0.0152 \\
SWIFT J2232.4+0814  & Ark 557                 & 14.84 & 50.13 & $1.84 \times 10^{44}$  & 0.0252 \\
SWIFT J0202.4+6824A & 2MASX J02021731+6821460 & 14.72 & 44.10 & $2.94 \times 10^{43}$  & 0.0119 \\
SWIFT J1212.9+0702  & NGC 4180                & 12.90 & 47.23 & $1.53 \times 10^{43}$  & 0.0070 \\
SWIFT J1001.7+5543  & NGC 3079                & 12.87 & 32.30 & $1.62 \times 10^{42}$  & 0.0037 \\
SWIFT J0423.5+0414  & 2MASX J04234080+0408017 & 10.06 & 42.09 & $6.37 \times 10^{44}$  & 0.0450 \\
SWIFT J2330.5-0228  & MCG-01-59-027           & 8.76  & 27.61 & $3.56 \times 10^{44}$  & 0.0334 \\
SWIFT J0744.0+2914  & UGC 03995A              & 7.76  & 29.27 & $2.21 \times 10^{43}$  & 0.0159 \\
\end{tabular}
\end{ruledtabular}
\end{table*}

We obtain the test statistic (TS) for each source as:
\begin{equation}
	TS = 2 \ln \frac{\mathcal{L}(\hat{n}_s, \hat{\gamma})}{\mathcal{L}(n_s=0)},
\end{equation}
where $\hat{n}_s$ and $\hat{\gamma}$ are the best-fit parameters. The TS value quantifies the significance of the neutrino excess at each source position. For a sufficiently large number of events, according to the Wilks' theorem, TS would asymptotically follow a $\chi^2$ distribution with two degrees of freedom under the null hypothesis (corresponding to the two free parameters $n_s$ and $\gamma$), and the local significance is therefore derived from the cumulative distribution function of the $\chi^2_{\rm dof=2}$ distribution.
It should be noted that the two parameters $n_s$ and $\gamma$ are correlated, which may effectively reduce the degrees of freedom and lead to deviations from the $\chi^2_{\rm dof=2}$ expectation. However, for the purposes of this study, an exact significance value is not essential, and we therefore retain the $\chi^2_{\rm dof=2}$ distribution as an indicative estimate.

The neutrino flux normalization at 1 TeV, $\Phi_0$ (in units of TeV$^{-1}$ cm$^{-2}$ s$^{-1}$), is obtained from the maximum likelihood fit assuming a power-law spectrum $dN/dE = \Phi_0 (E/1\,{\rm TeV})^{-\gamma}$. The energy-integrated neutrino flux (for a single flavor, $\nu_\mu$) is:
\begin{equation}
	F_\nu = \int_{E_{\rm min}}^{E_{\rm max}} \Phi_0 \left(\frac{E}{1\,{\rm TeV}}\right)^{-\gamma}E dE,
\end{equation}
where $E_{\rm min} = 0.3$ TeV and $E_{\rm max} = 100$ TeV.
The neutrino luminosity is then computed as: $L_\nu = 4\pi D_L^2 \cdot 3F_\nu$,
where the factor of $4\pi$ accounts for the assumption of isotropic emission, and the factor of 3 accounts for flavor equipartition at Earth ($\nu_e:\nu_\mu:\nu_\tau \approx 1:1:1$ due to neutrino oscillations over cosmic distances).

\section{Results}
\label{sec:results}

\begin{figure*}[t]
	\centering
	\includegraphics[width=0.40\textwidth]{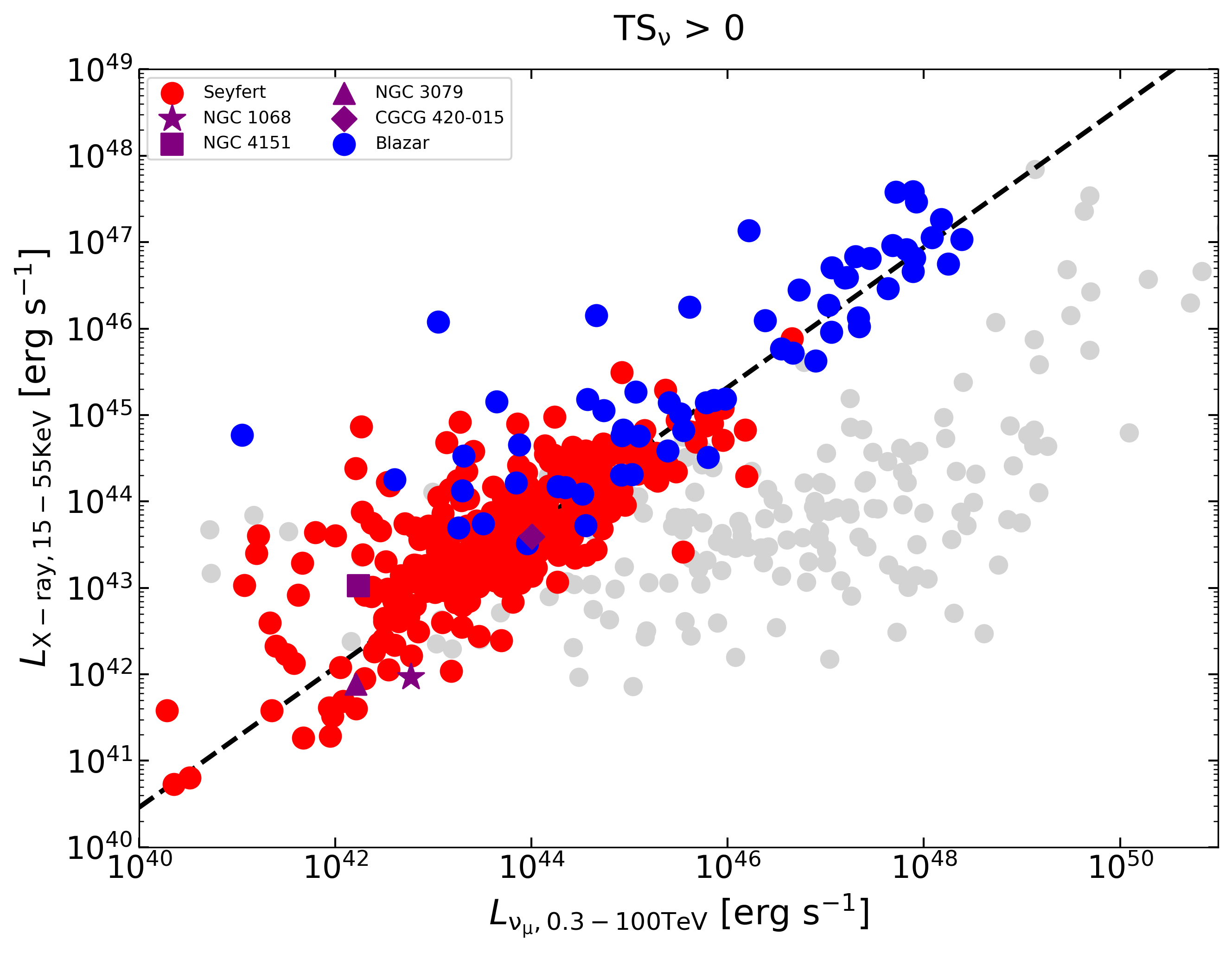}
	\includegraphics[width=0.40\textwidth]{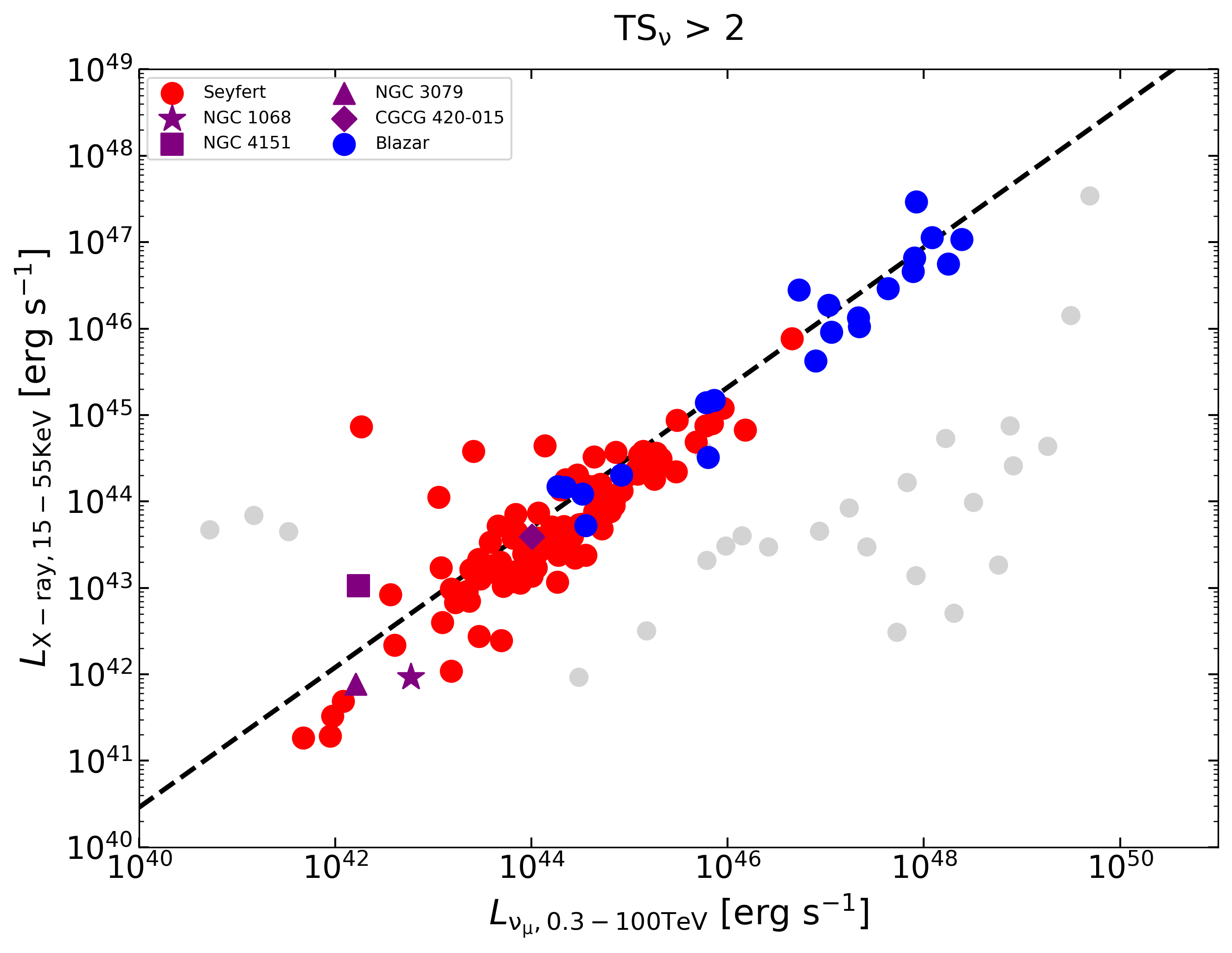}\\
	\includegraphics[width=0.40\textwidth]{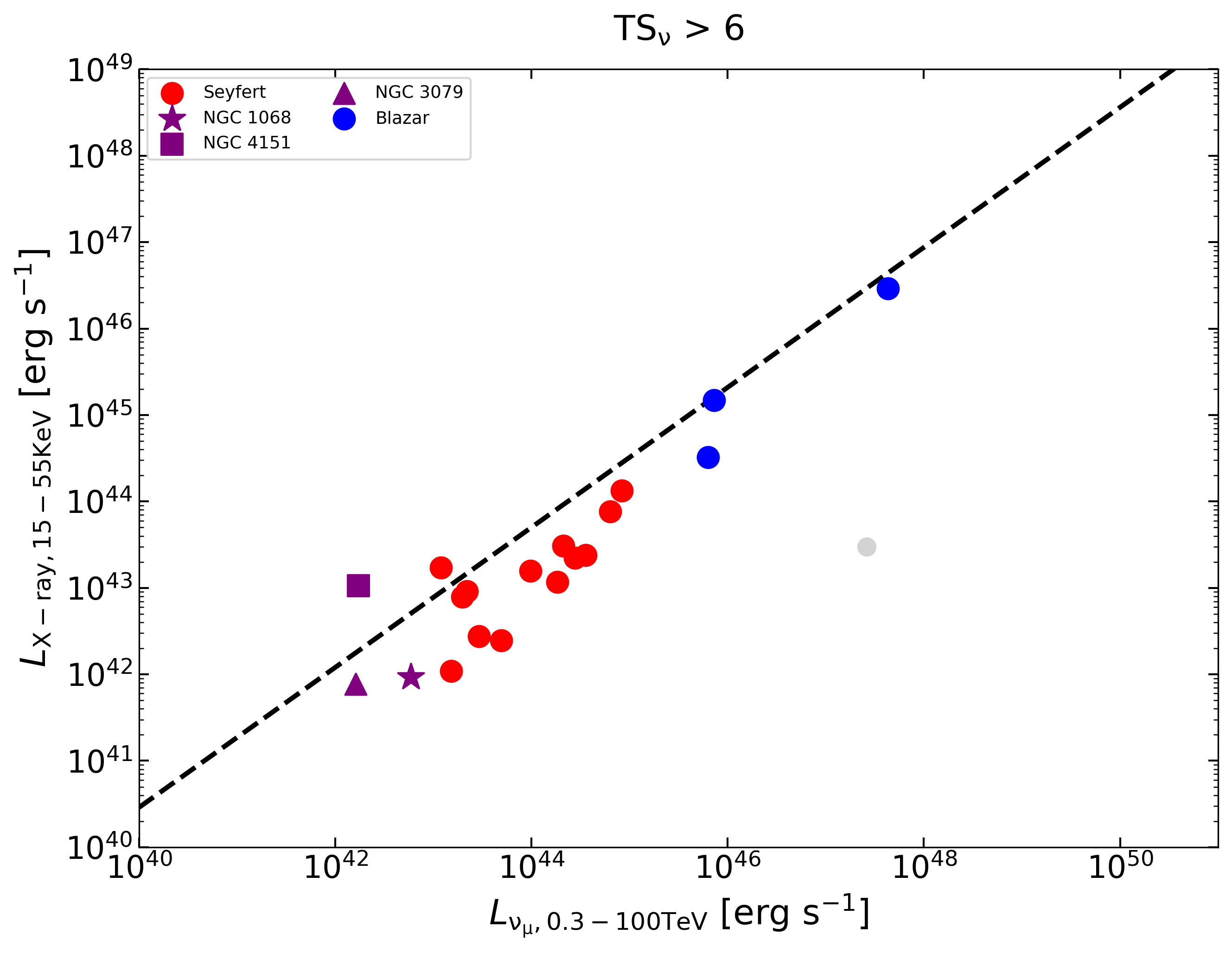}
	\includegraphics[width=0.40\textwidth]{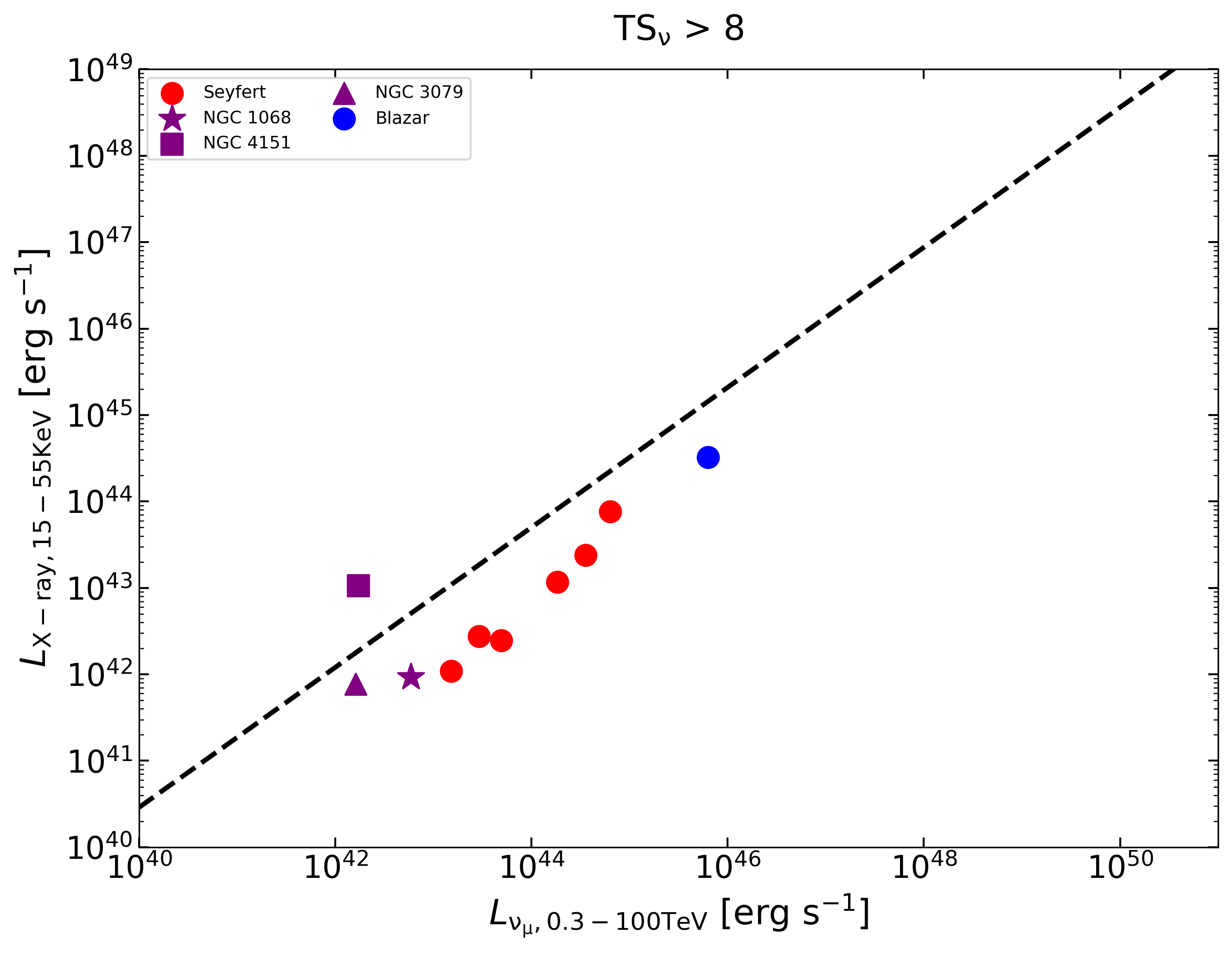}
	\caption{Observed apparent correlation between neutrino luminosity and hard X-ray luminosity ($L_\nu$--$L_X$) for the AGN sample under different TS cuts. {Panels (left to right, top to bottom):} TS $> 0$, TS $> 2$, TS $> 6$, and TS $> 8$. The neutrino luminosity $L_\nu$ is derived from the likelihood analysis of the IceCube 10-yr public muon-track data. The X-ray luminosity is derived using the X-ray flux reported in the Swift AGN catalog. Red and blue markers denote Seyfert galaxies and blazars, respectively; gray points indicate Southern sky sources ($b<-10^\circ$), which are excluded due to the reduced sensitivity of IceCube's track-like data in the Southern hemisphere. As the TS threshold increases, the number of sources decreases but all panels show a pronounced correlation.}
	\label{fig:obs_correlation}
\end{figure*}

\begin{figure*}[t]
	\centering
	\includegraphics[width=0.45\textwidth]{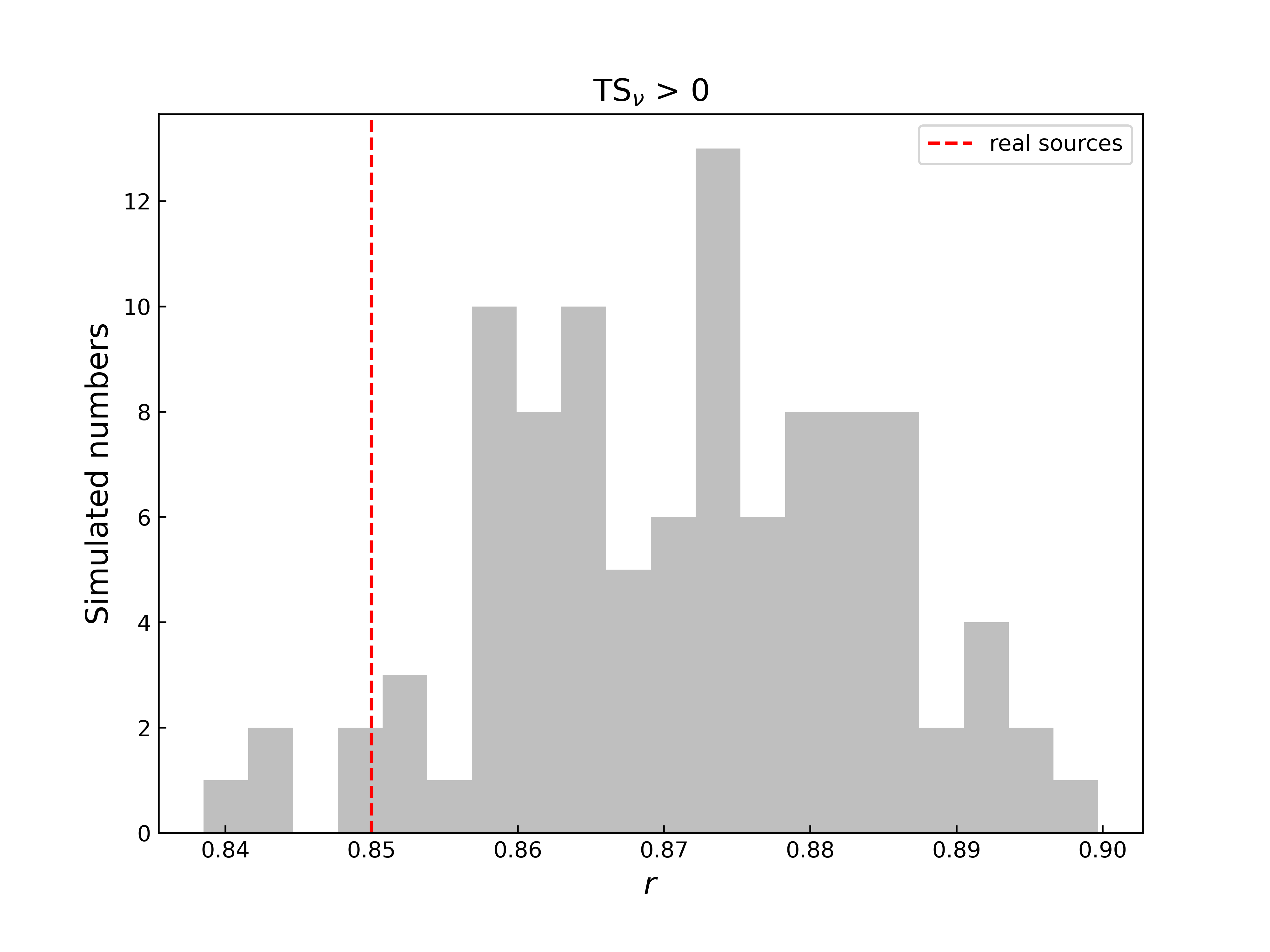}
	\includegraphics[width=0.45\textwidth]{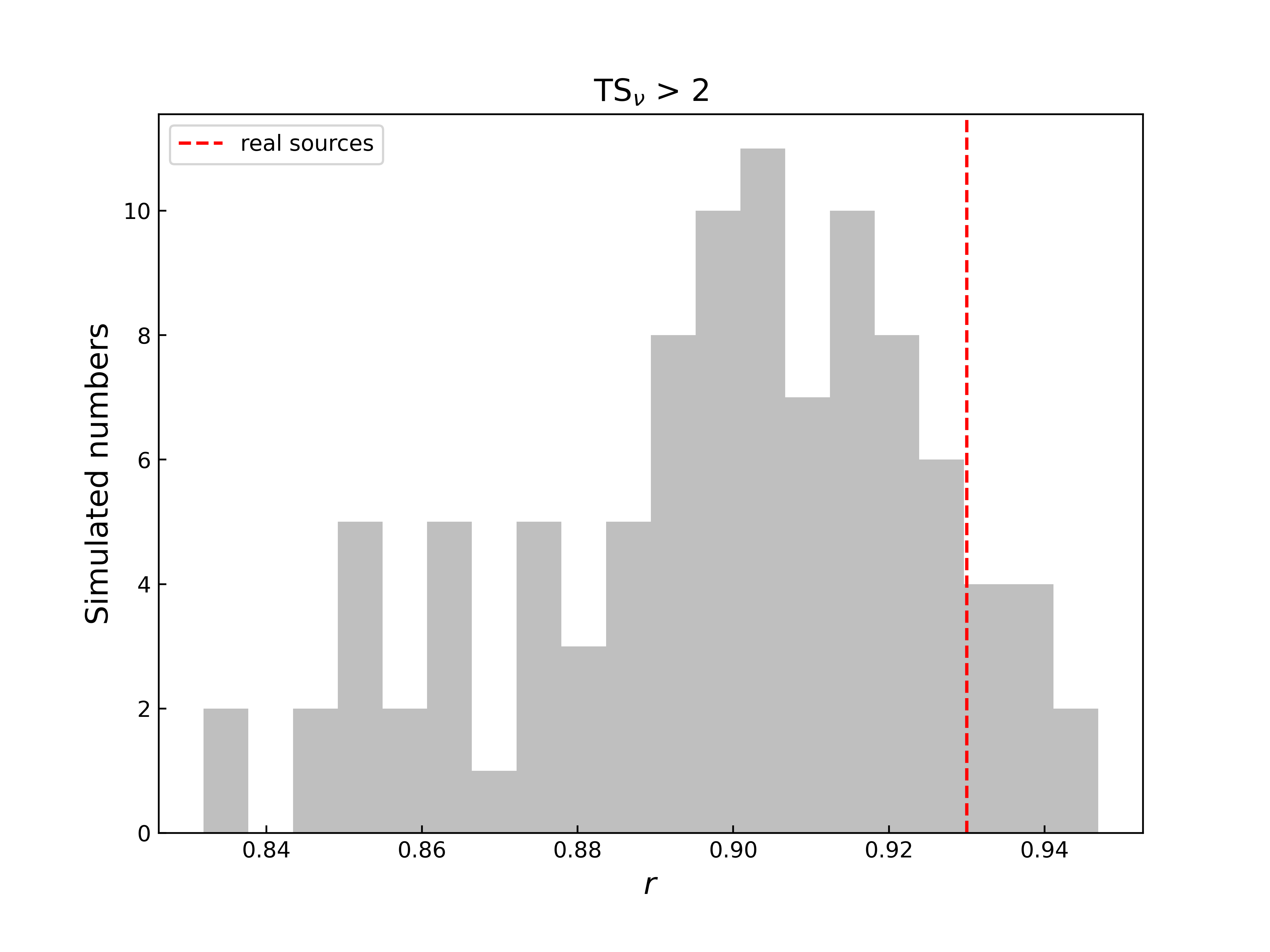}\\
	\includegraphics[width=0.45\textwidth]{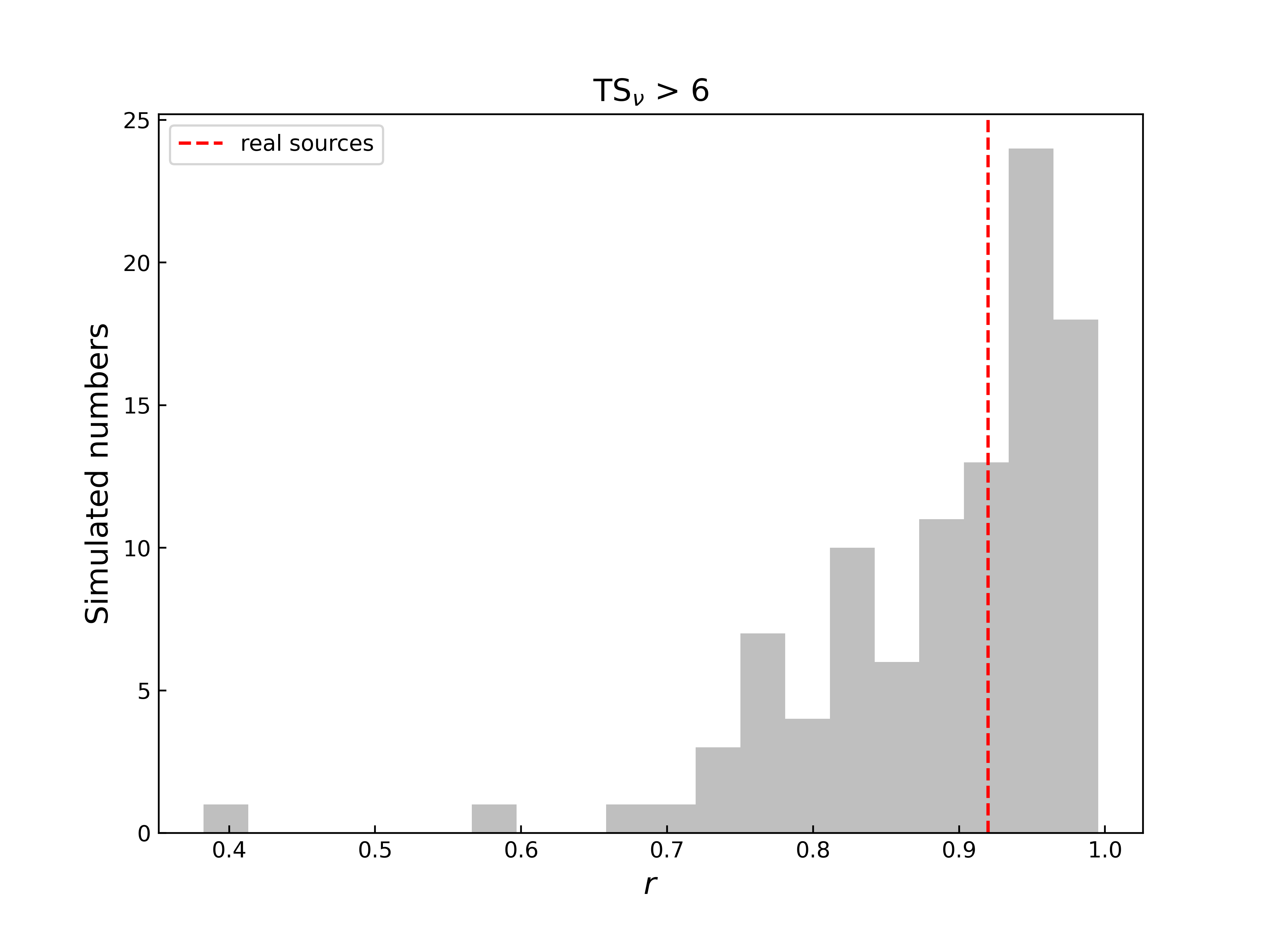}
	\includegraphics[width=0.45\textwidth]{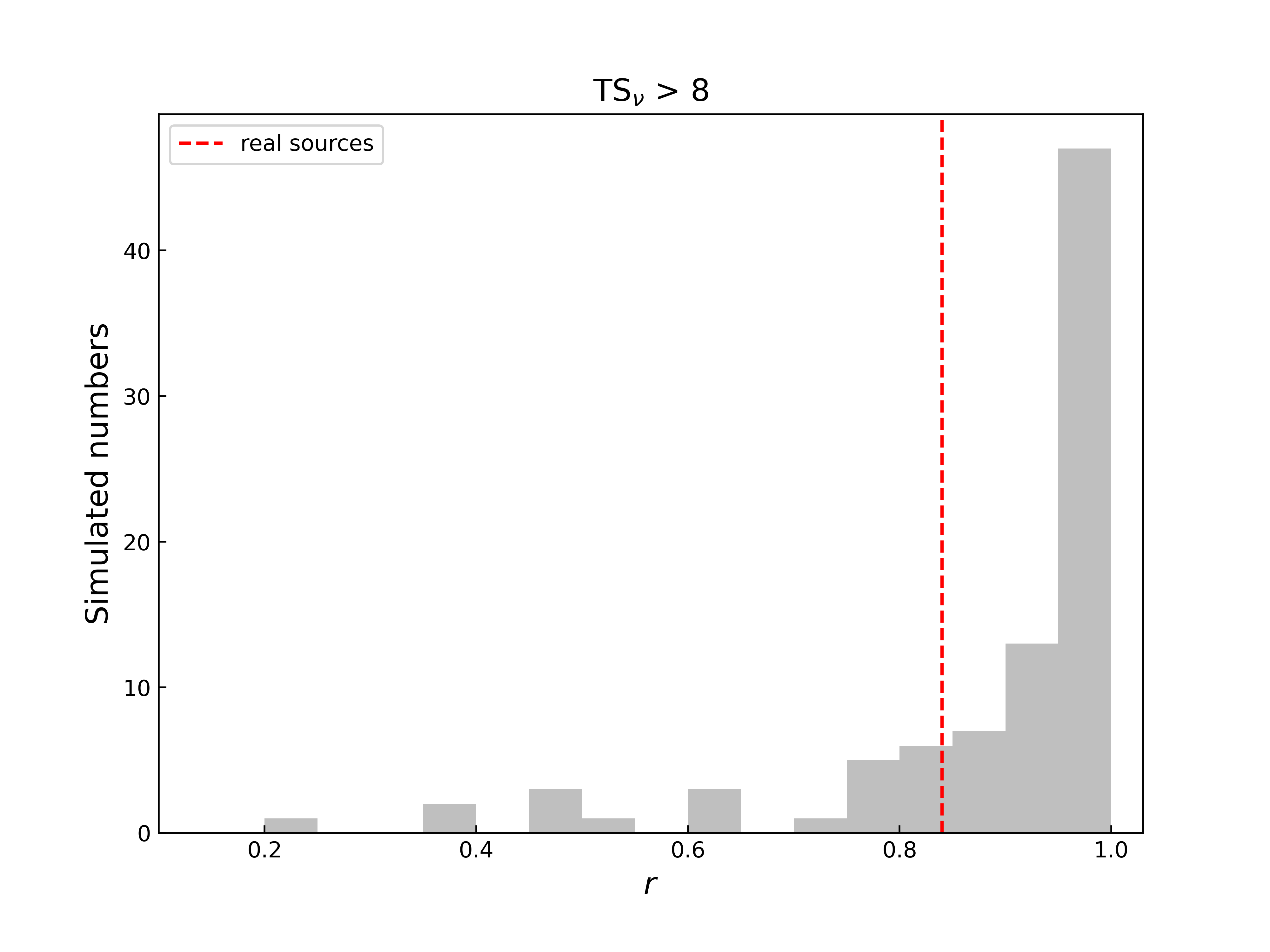}
	\caption{Distribution of the Pearson correlation coefficient $r$ derived from 100 simulated samples, each containing random sky positions with no physical association to the X-ray catalog. In each simulation, the right ascensions are uniformly randomized and constrained to match the original distribution via KS tests, while the declinations and X-ray properties are preserved from the real sample. Four panels correspond to different TS cuts. The dashed red line marks the $r$ value of the real sources. The obtained $r$ is usually high ($r\gtrsim0.8$), despite the absence of real sources at the random positions.}
	\label{fig:sim_r_P_L}
\end{figure*}

\begin{figure*}[t]
	\centering
	\includegraphics[width=0.40\textwidth]{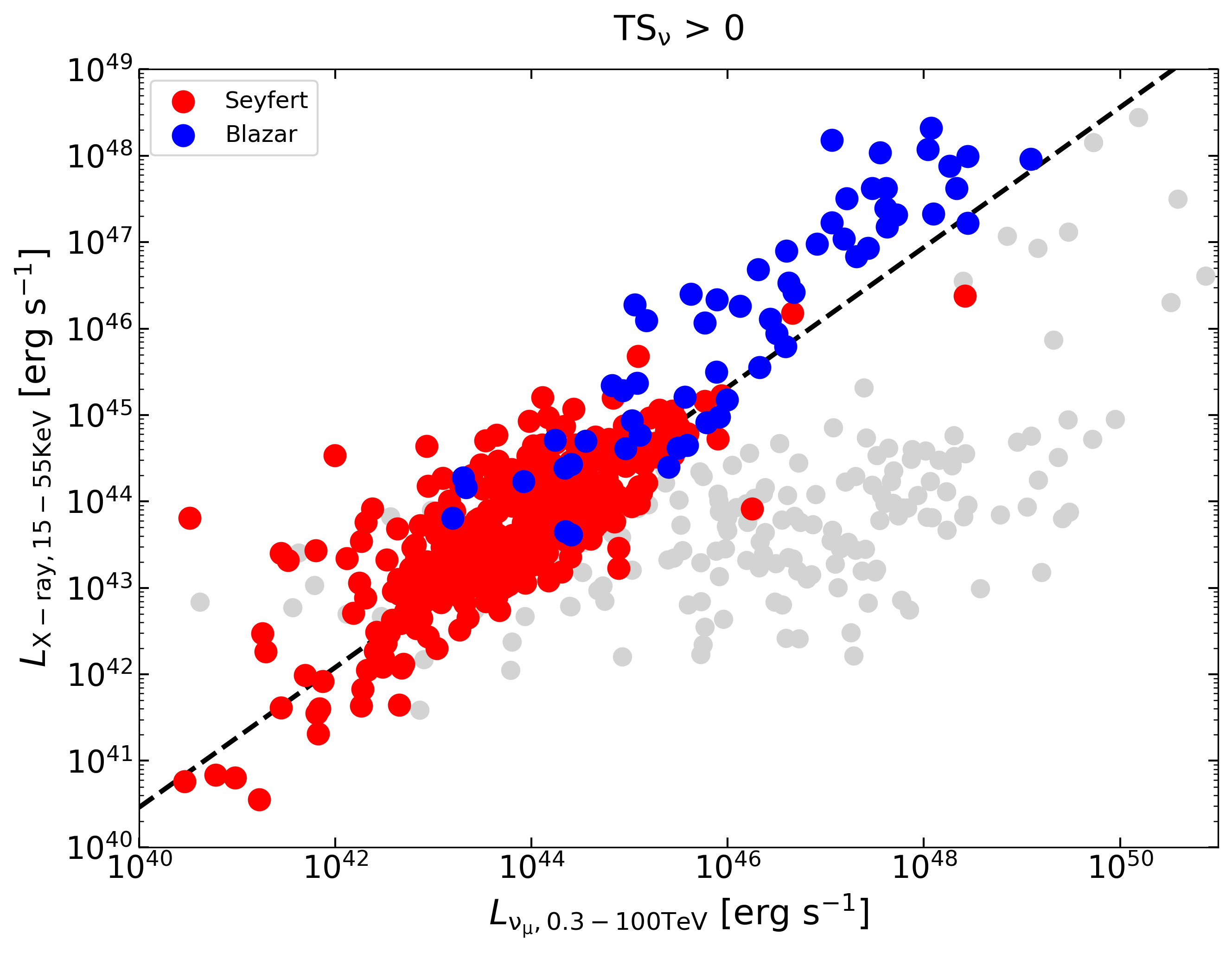}
	\includegraphics[width=0.40\textwidth]{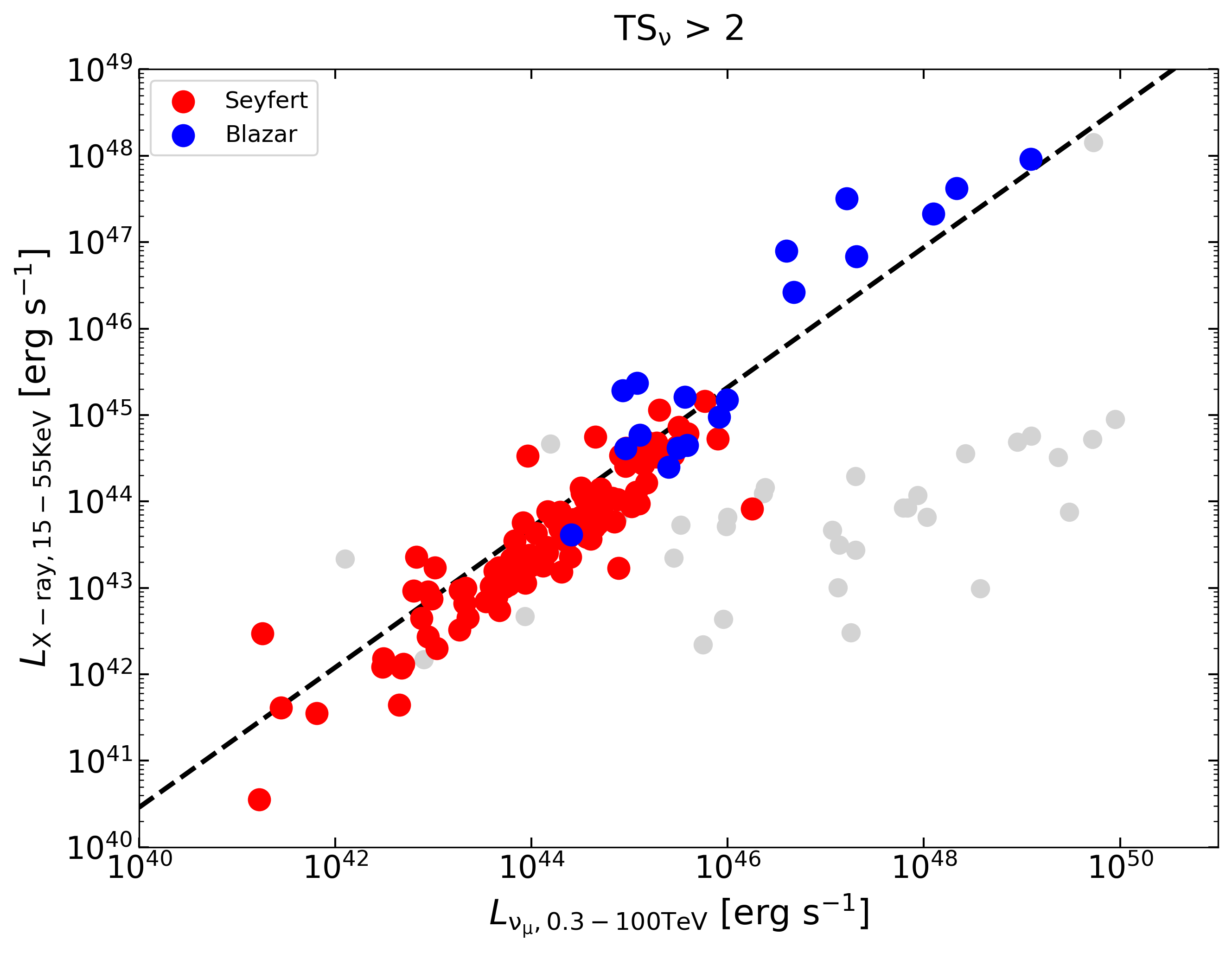}\\
	\includegraphics[width=0.40\textwidth]{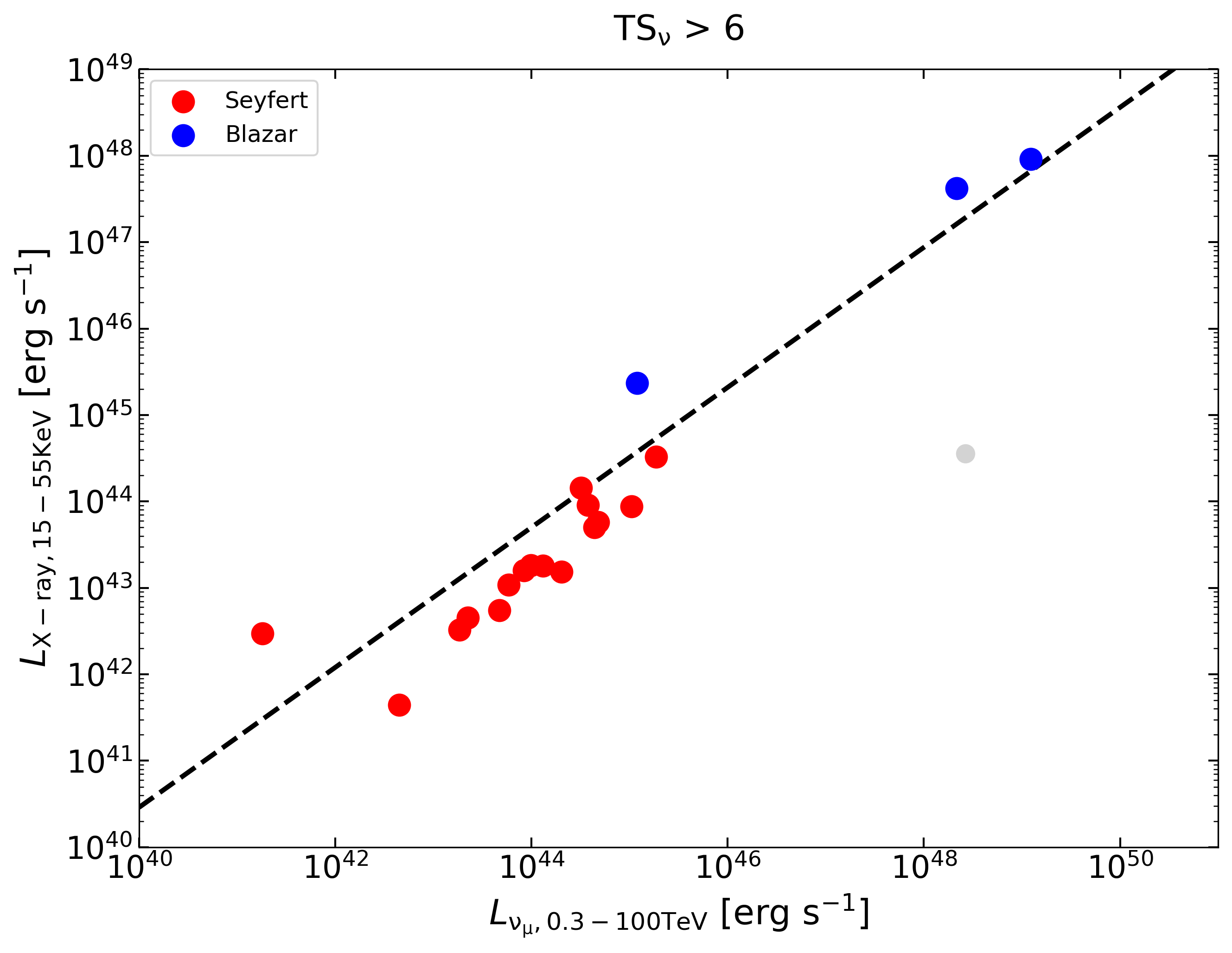}
	\includegraphics[width=0.40\textwidth]{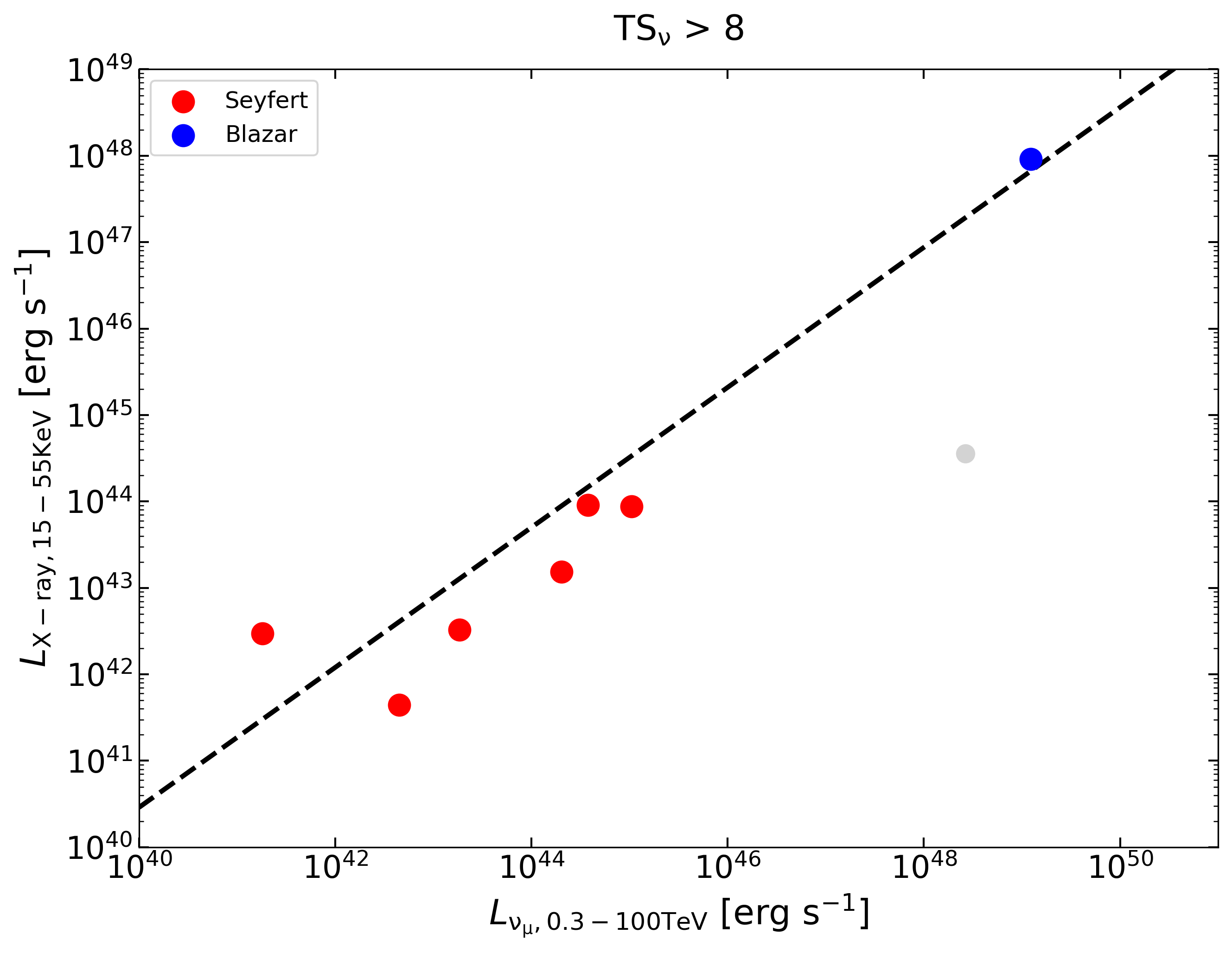}
	\caption{One example of $L_\nu$--$L_X$ relation from simulated sample. Four panels correspond to different TS cuts. Despite the absence of real sources at the random sky positions, the simulated data reproduces the same apparent linear correlation as the real data.}
	\label{fig:sim_correlation}
\end{figure*}

\subsection{Real Observation}

Using the method described above, we obtained the X-ray and neutrino luminosities for all 1114 sources in our sample.
{Table~\ref{tab:top10} lists the 10 sources with the highest TS values from the neutrino analysis of the AGN sample. The two most significant detections correspond to the well-established neutrino sources NGC 1068 (TS $\approx 19.8$) and NGC 4151 (TS $\approx 18.4$), both of which have been reported in previous IceCube analyses \citep{IceCube:2024ayt,Abbasi2024a,IceCube:2025gdd}.
The TS value for NGC 1068 is slightly lower than that reported by the IceCube Collaboration, which is due to the shorter observation period of the public data and the fixed-coordinate analysis (i.e., evaluating the likelihood at the catalog position rather than scanning the local region to maximize the TS).
The remaining sources in the top 10 exhibit TS values in the range of $\sim$8--15, corresponding to local significances of $\sim$2.4--3.5$\sigma$.}

Fig.~\ref{fig:obs_correlation} shows the $L_\nu$--$L_X$ plane for our AGN sample with different TS cuts. Red points represent Seyfert galaxies and blue points represent blazars. Gray points represent sources in the Southern sky; due to IceCube's significantly lower sensitivity to the Southern hemisphere for muon track data, current point source candidates are all from the Northern sky, and here we focus on Northern hemisphere sources. The four neutrino Seyfert galaxy candidates NGC 1068, NGC 4151, NGC 3079, and CGCG 420-015 are shown with different markers in the figure. The neutrino blazar candidates TXS 0506+056 and PKS 1424+240 are not listed in the \textit{Swift}-BAT 157-month catalog because their hard X-ray fluxes lie below the catalog's sensitivity limit, and are therefore absent from our sample.

For TS $> 0$ (top left panel), the sample includes 589 sources (other sources have TS $= 0$ or slightly negative values due to numerical effects), and the points are widely scattered with no obvious correlation. However, when we restrict the sample to Northern hemisphere sources (excluding the Southern sky with $b < -10^\circ$), the remaining points begin to exhibit a clear correlation. As we increase the TS threshold to 2 (top right panel), 6 (bottom left panel), and 8 (bottom right panel), the number of sources decreases, but the remaining points keep clustering around the linear relation in log-log space.

The linear correlation between neutrino and X-ray luminosities can be quantified using the Pearson correlation coefficient $r$. In log-luminosity space, the samples with different TS threshold cuts yield $r$ values of 0.85, 0.93, 0.92, and 0.84, respectively. These results are consistent with previous studies \citep{Kun2024,Kun2025}, which also present a similarly strong linear trend.

If interpreted directly, this progression might suggest a real correlation because the correlation becomes clearer as we select more significant neutrino candidates. However, as demonstrated by the Monte Carlo simulations below, an identical behavior emerges for random sky positions, indicating that the trend is driven by selection effects rather than an intrinsic physical connection.

\subsection{Monte Carlo Simulation}
\label{MC}
We next perform Monte Carlo simulations using fake source positions that are not associated with any real astrophysical objects. We generate 964 and 150 random sky positions corresponding to the numbers of Seyfert galaxies and blazars in our actual sample, respectively. For each source in the sample, we randomize its right ascension uniformly over the range $0^\circ$--$360^\circ$, while preserving its original declination. 
The simulated right ascensions are further constrained to match the original distribution using the Kolmogorov-Smirnov (KS) test, following the procedure described in \citet{Lu:2024flp}.
This procedure ensures that the declination distribution of the simulated sample exactly matches that of the real sample, thereby properly accounting for IceCube's declination-dependent sky exposure, which arises from the detector's geographical position at the South Pole and its anisotropic sensitivity to through-going muon tracks across the Northern and Southern hemispheres.

For each random position, we conduct the identical point-source likelihood analysis that we apply to the real AGN sample using the same IceCube dataset. Since these positions contain no genuine astrophysical sources, any excess ``signal'' detected represents purely statistical background fluctuations. The TS values and flux normalizations extracted from this analysis therefore characterize the noise properties of the measurement rather than physical emission.

Following the neutrino analysis, we assign to each random position the redshift and X-ray flux from the real BAT catalog source.
We then compute fake neutrino luminosities using the neutrino flux normalizations obtained from the random positions combined with the luminosity distances corresponding to the X-ray properties, employing the same conversion formula applied to the real data.

Finally, we construct $L_\nu$--$L_X$ diagrams for the simulated dataset using identical TS thresholds as employed for the real data. This simulation procedure effectively decouples any physical connection between neutrino and X-ray emission while fully preserving the statistical properties of the analysis and the selection effects introduced by the TS cuts. Any correlation appearing in the simulated data must therefore arise solely from these methodological artifacts rather than from intrinsic source physics.

Fig.~\ref{fig:sim_r_P_L} shows the distributions of $r$ derived from 100 simulated samples (random sky positions), while a representative $L_\nu$--$L_X$ correlation from an individual simulation is illustrated in
Fig.~\ref{fig:sim_correlation}. Due to the absence of real sources, a physical connection between the ``neutrino'' signals (which are pure background fluctuations) and the X-ray properties is not expected. However, the simulated data reproduces the same apparent correlation as the real data.
This result strongly suggests that the correlation observed in the real data is not of physical origin.

\begin{figure*}[t]
	\centering
	\includegraphics[width=0.9\textwidth]{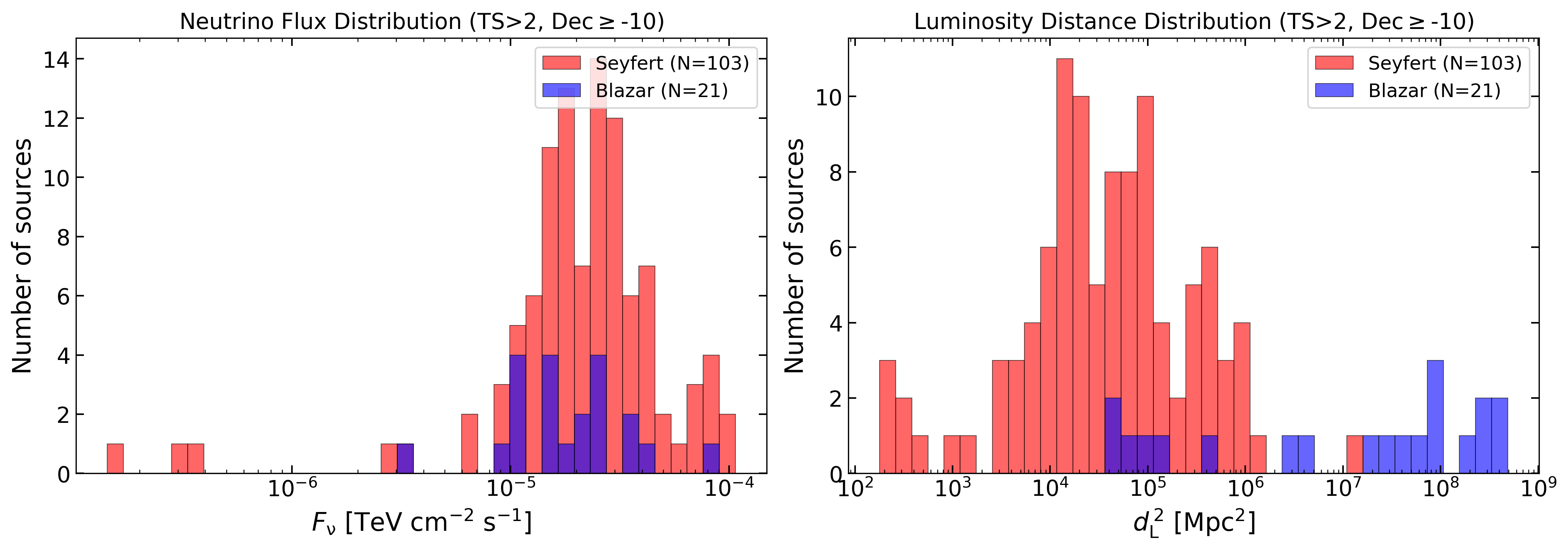}
	\caption{
	Left: Distribution of neutrino fluxes for sources with TS$>2$. The fluxes are confined to a narrow range spanning only a factor of $\sim$5. Right: Distribution of the square of luminosity distances for the AGN sample, spanning $\sim$7 orders of magnitude. Because luminosity scales as $L \propto D_L^2 F$, the large $D_L^2$ range causes the luminosity to be dominated by distance rather than intrinsic flux.}
	\label{fig:flux_dist}
\end{figure*}

\begin{figure*}[t]
	\centering
	\includegraphics[width=0.45\textwidth]{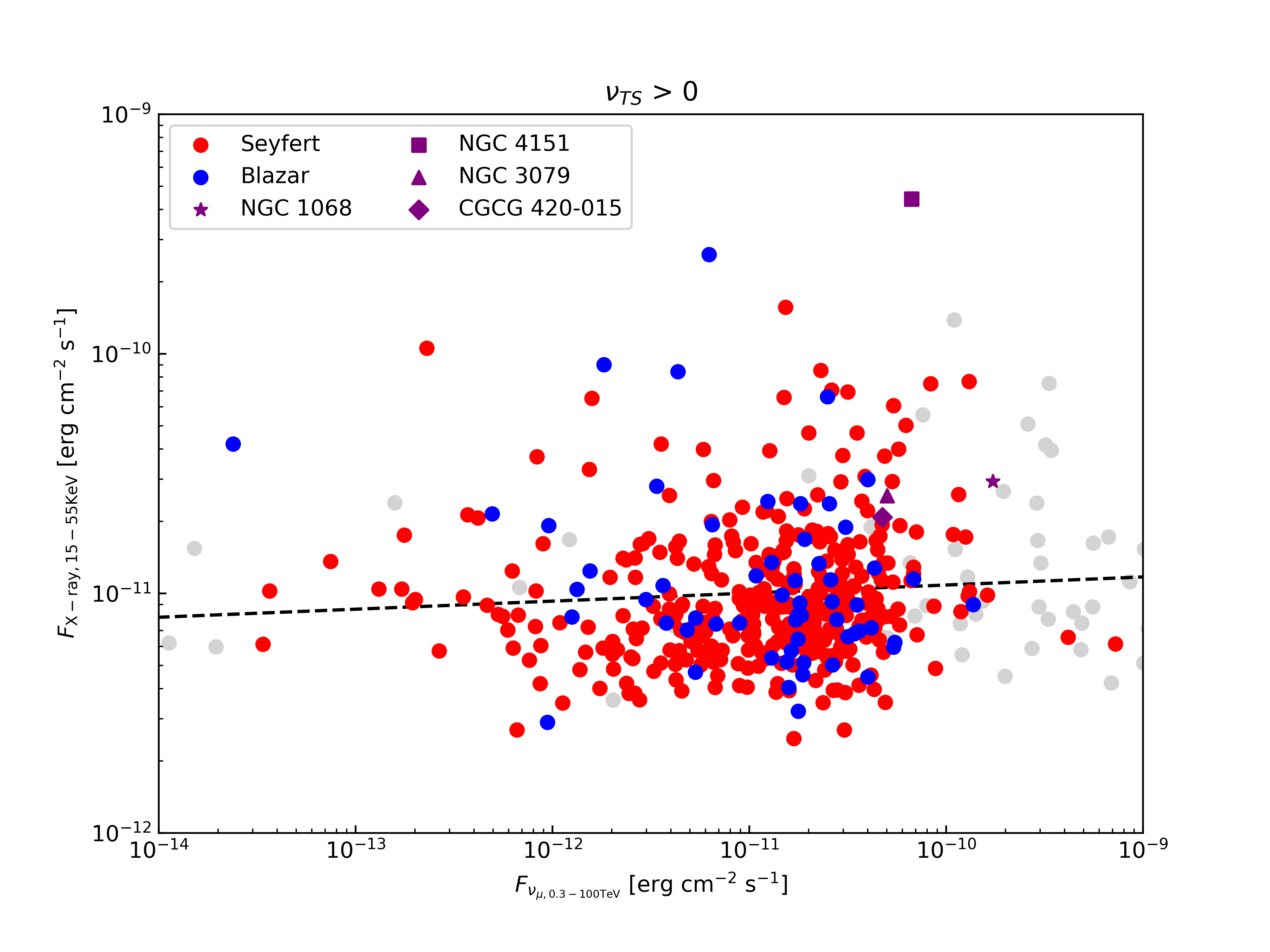}
	\includegraphics[width=0.45\textwidth]{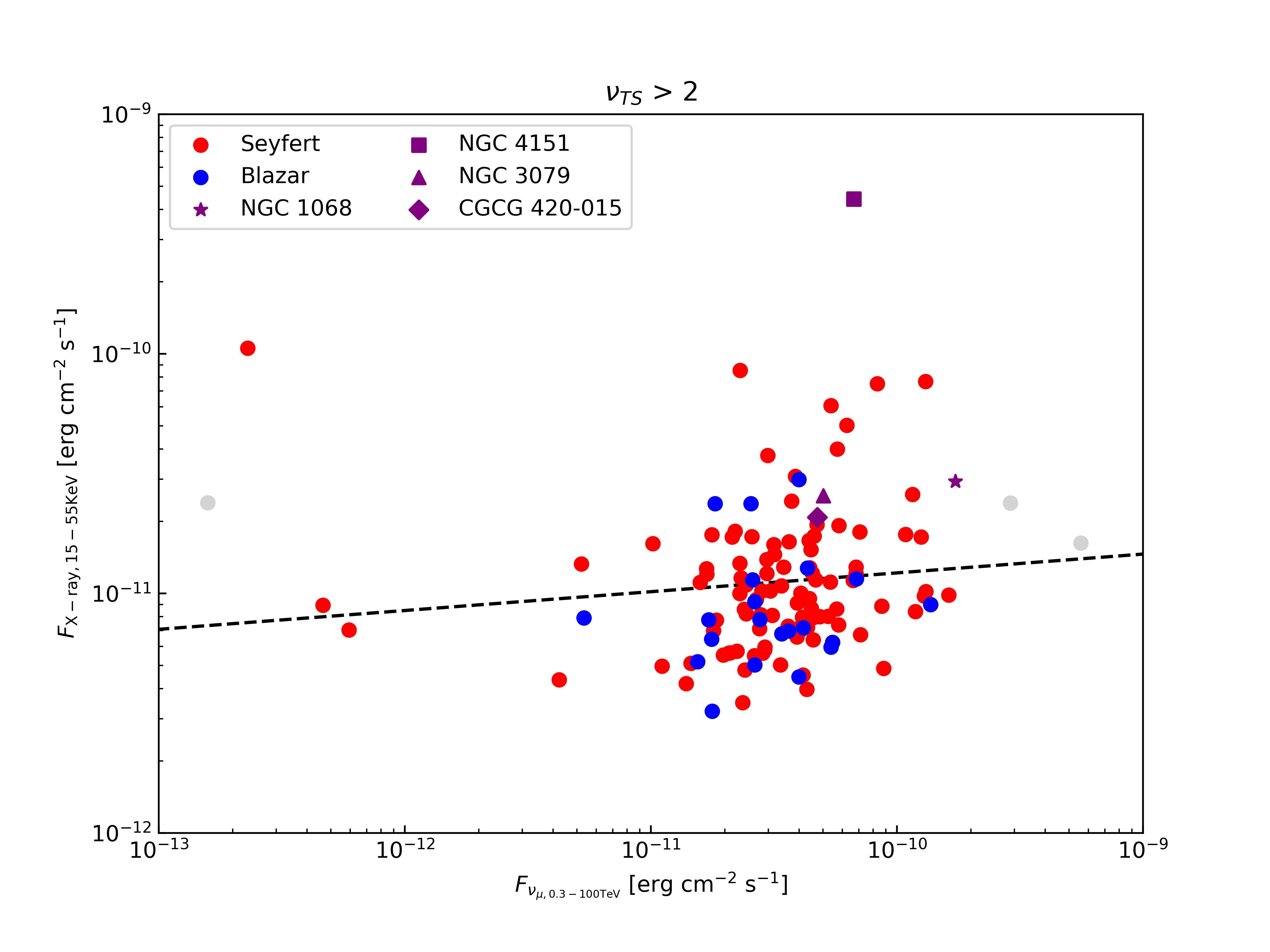}\\
	\includegraphics[width=0.45\textwidth]{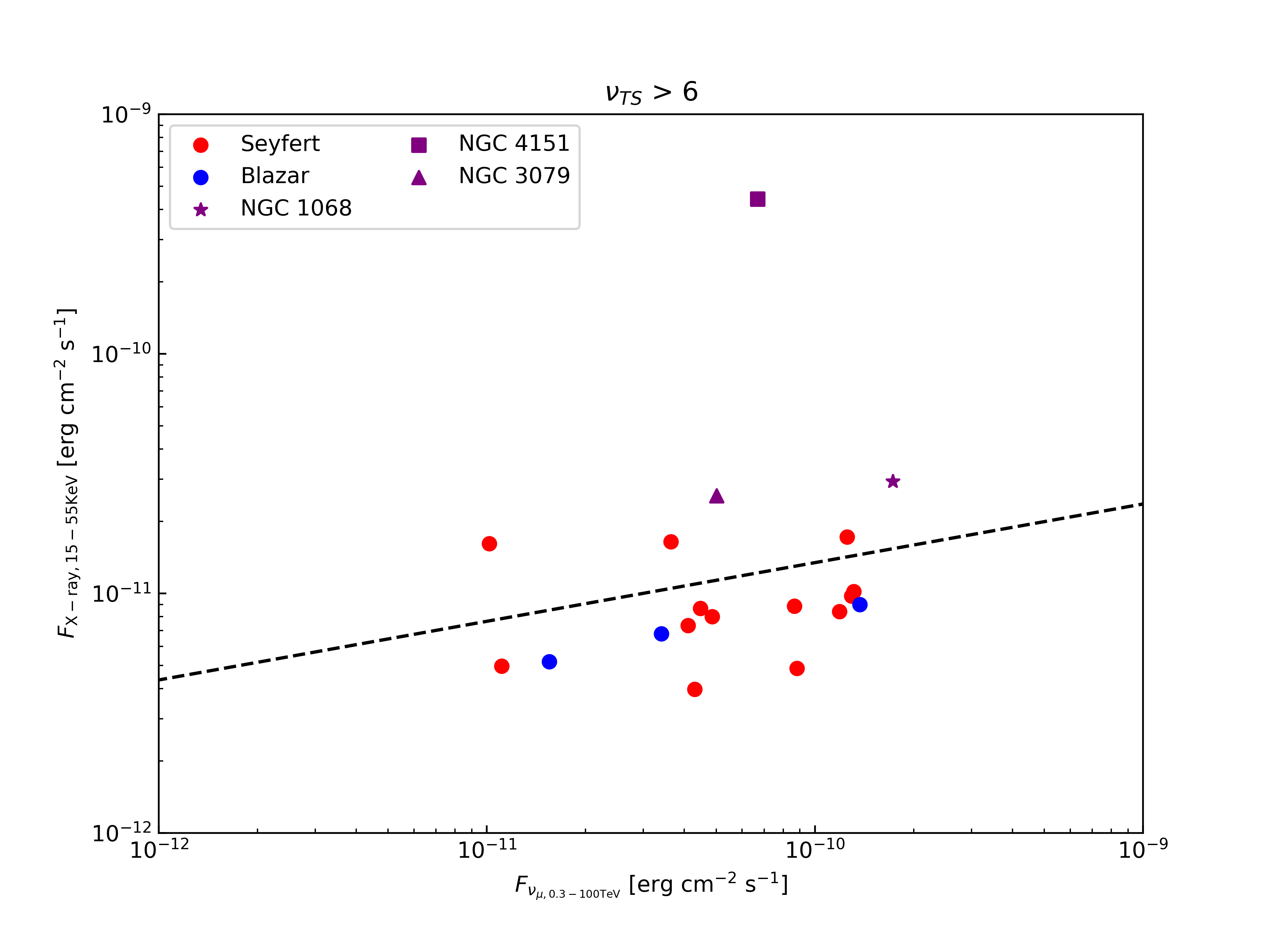}
	\includegraphics[width=0.45\textwidth]{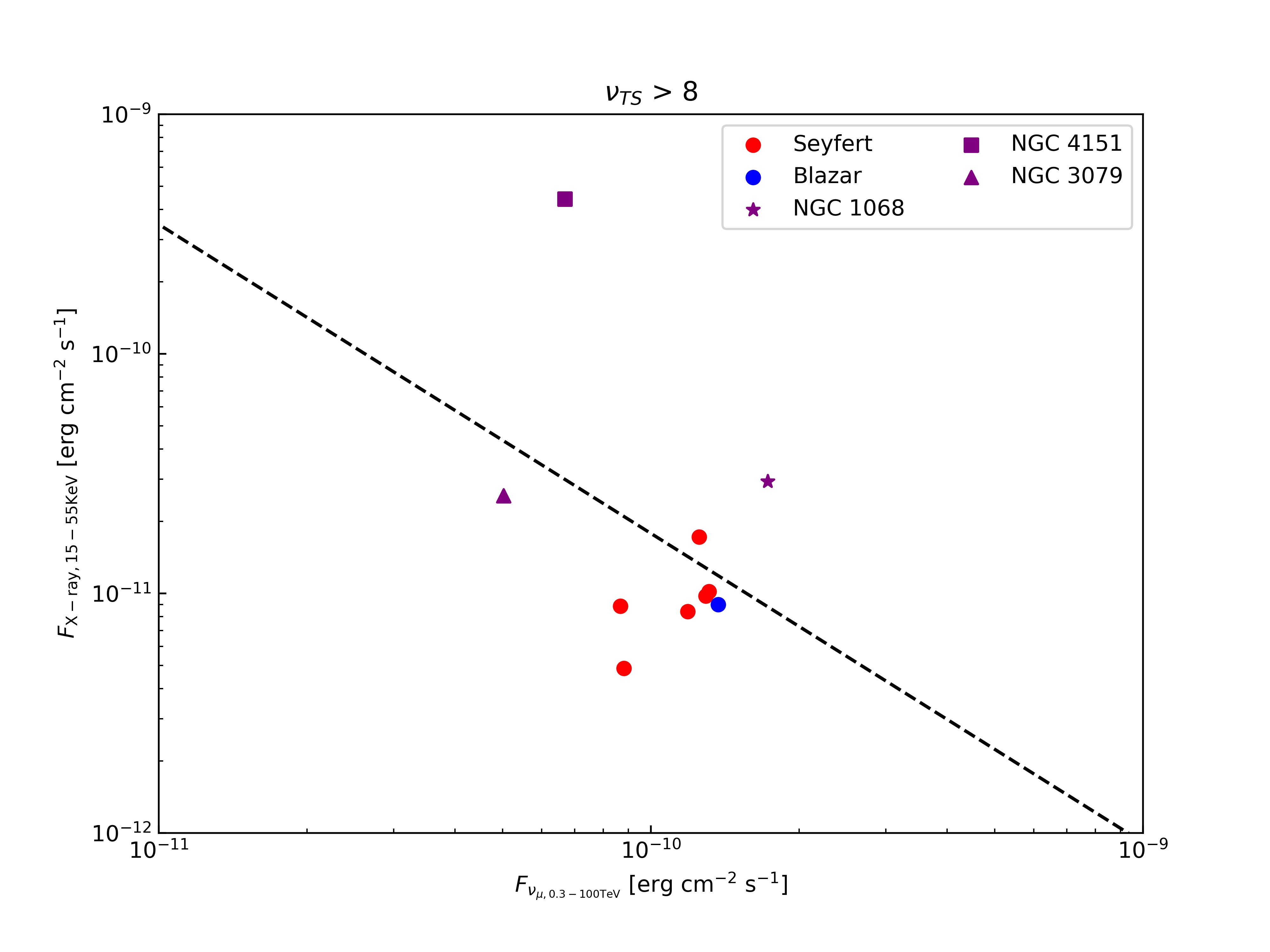}
	\caption{The same as Fig.~\ref{fig:obs_correlation}, but for the $F_\nu-F_X$ correlation.}
	\label{fig:obs_flux_correlation}
\end{figure*}

\begin{figure*}[t]
	\centering
	\includegraphics[width=0.45\textwidth]{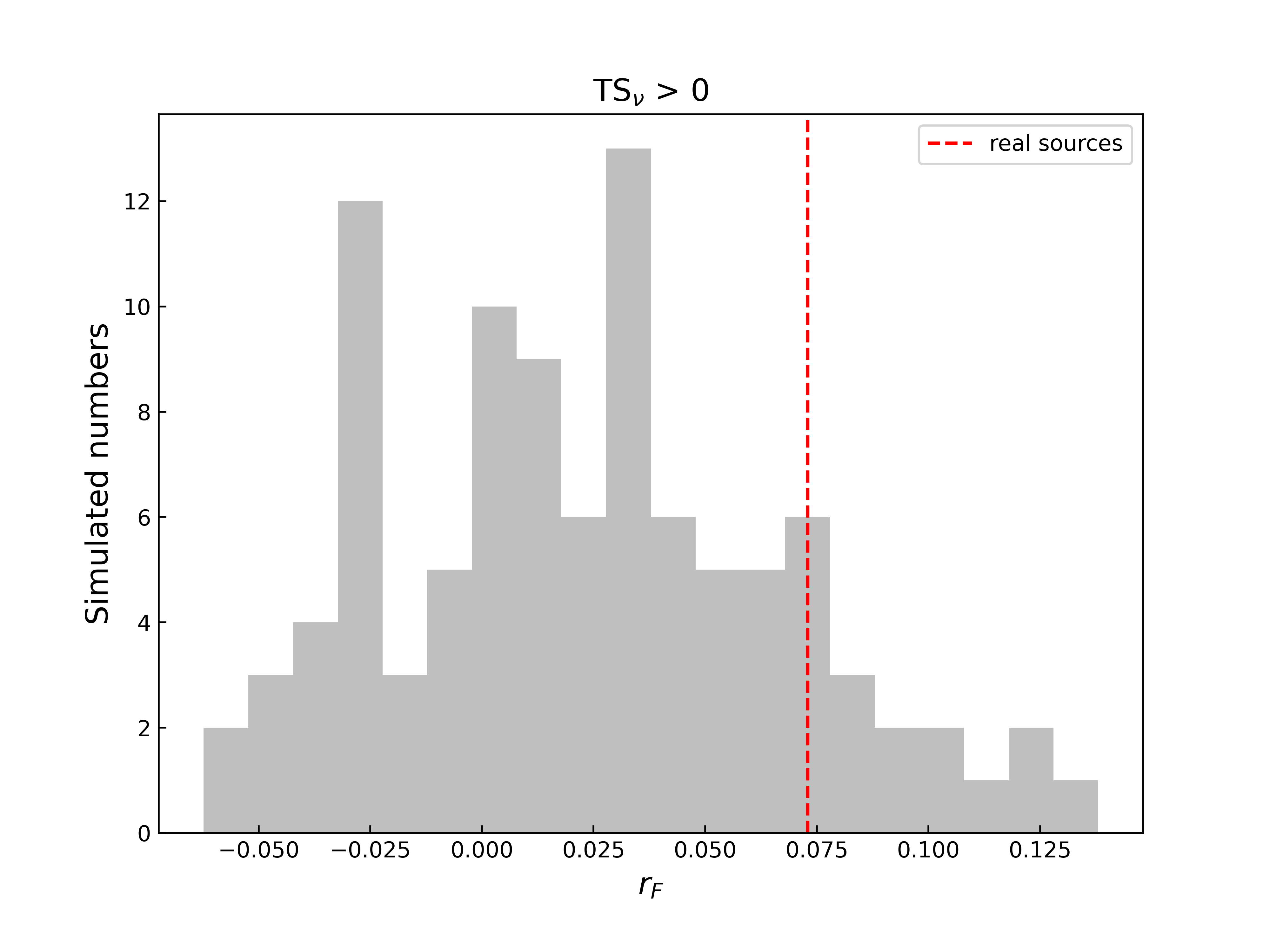}
	\includegraphics[width=0.45\textwidth]{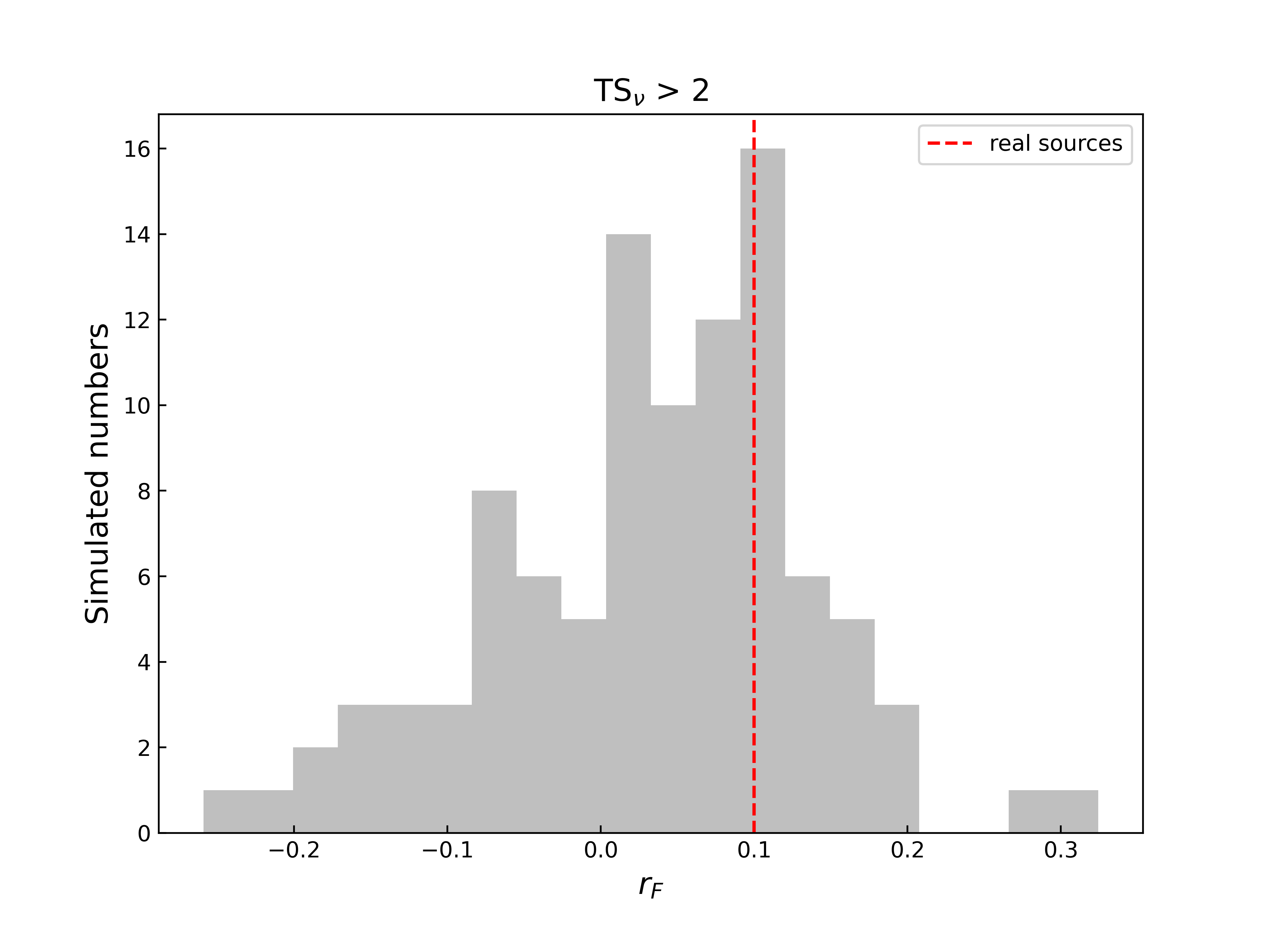}\\
	\includegraphics[width=0.45\textwidth]{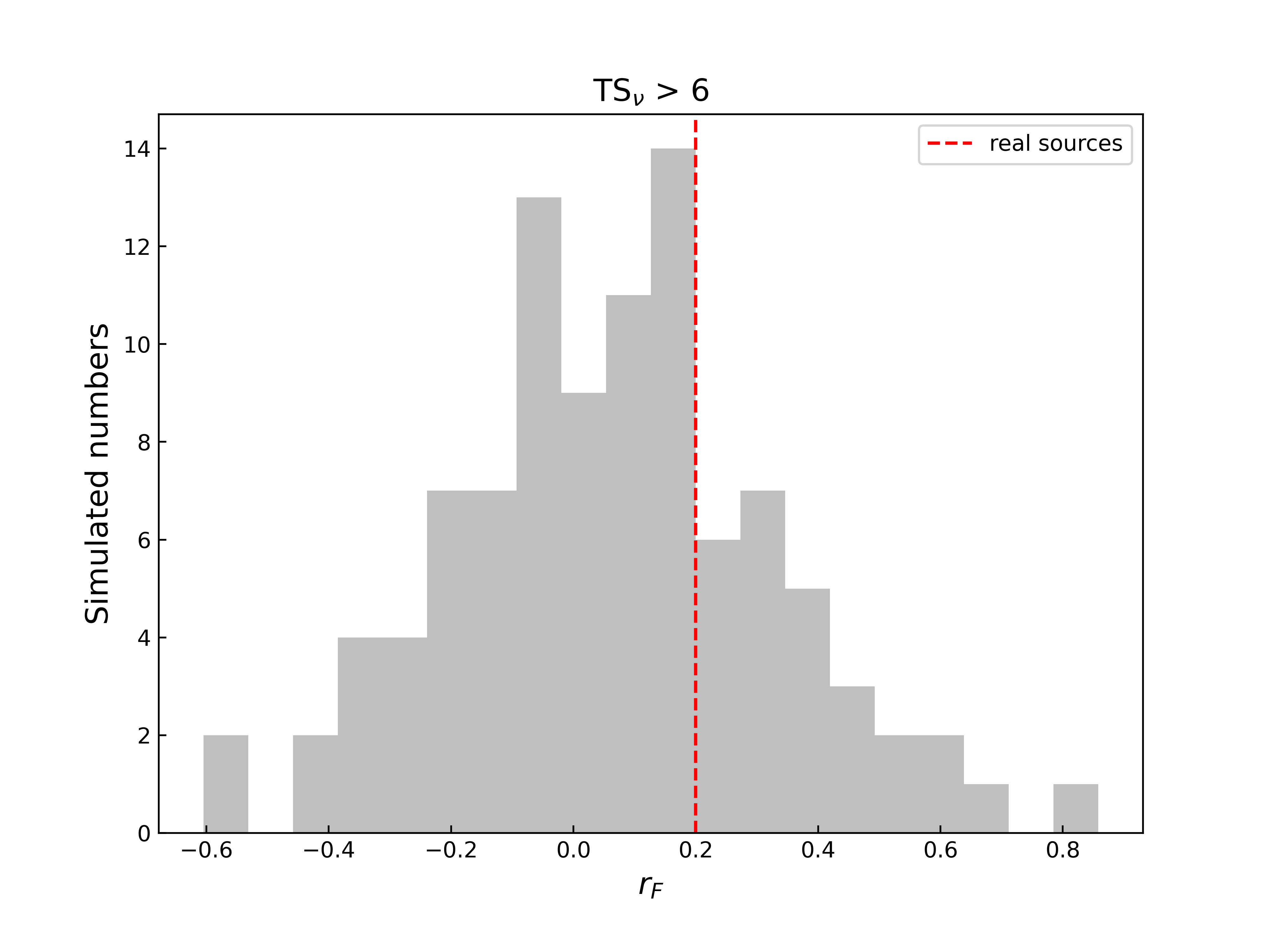}
	\includegraphics[width=0.45\textwidth]{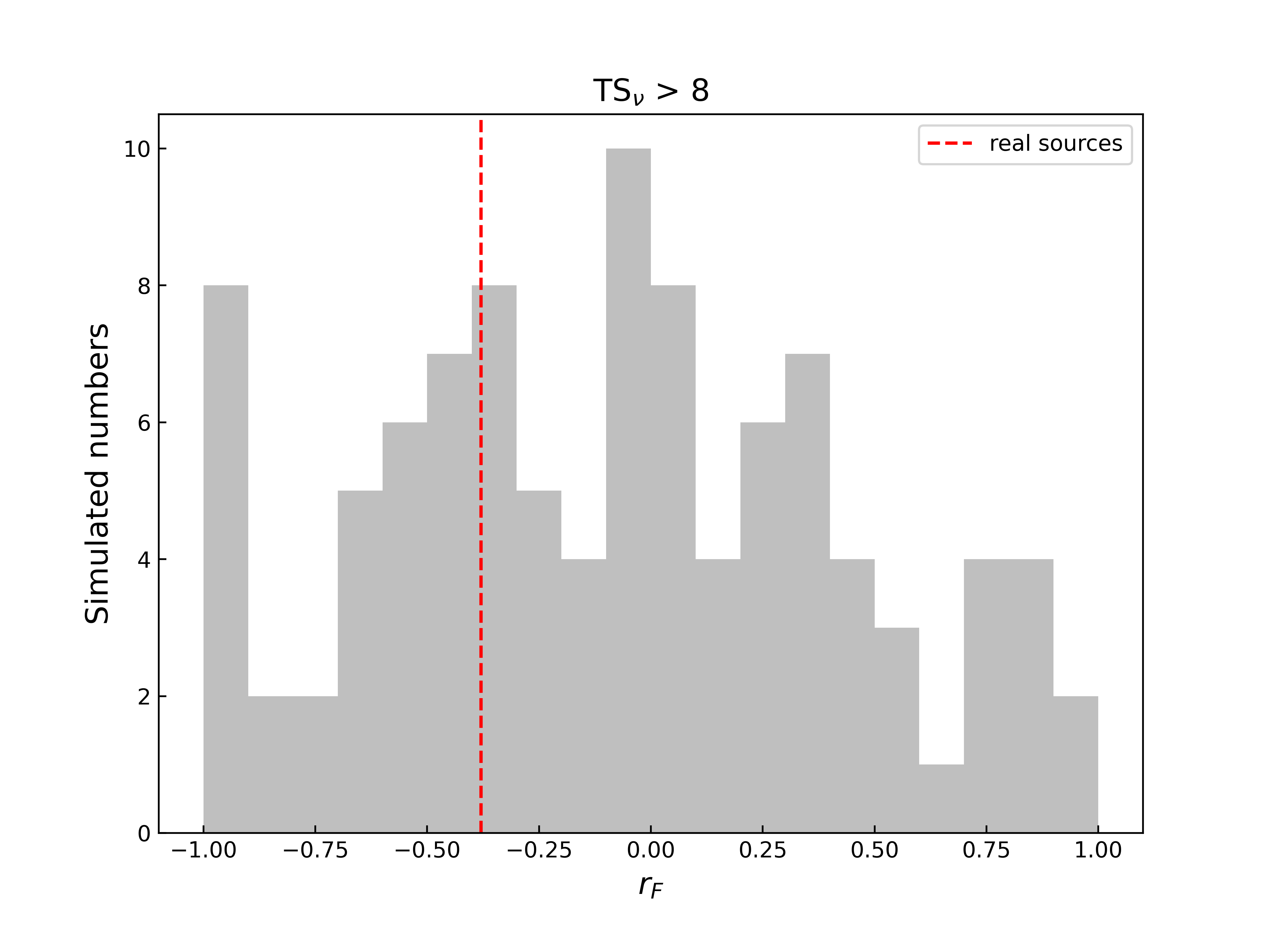}
	\caption{The same as Fig.~\ref{fig:sim_r_P_L}, but for the $r_F$ distribution. The distribution width broadens with the TS threshold, primarily due to the reduced sample size. Stricter cuts retain fewer sources, leading to larger statistical fluctuations in the correlation coefficient.}
	\label{fig:sim_r_P_F}
\end{figure*}
\subsection{Interpretation of the Apparent Correlation}

We next examine the origin of this apparent correlation.
The key issue is that the TS threshold (or p-value threshold or $\sigma$ threshold, they are convertible) effectively selects sources within a narrow range of neutrino fluxes. For a background-limited search like IceCube's point-source analysis, the TS value scales approximately as: $TS \approx {n_s^2}/{n_{\rm bg}}$,
where $n_s$ is the number of signal events and $n_{\rm bg}$ is the number of background events. For sources in similar background regions (which is approximately true for most of the Northern sky of IceCube data), a TS cut of TS $> 6$ (corresponding to $\sim$2.0$\sigma$) to TS $\sim$ 25 (corresponding to $\sim$4.5$\sigma$, roughly the largest significance for the IceCube point sources detected to date) corresponds to only a factor of $\sim$2--3 in flux (i.e., in $n_s$). 
From Table~\ref{tab:top10}, we can see that from TS$\sim8$ to 25 the best-fit signal event count ($n_s$) is confined within a relatively narrow factor of $\sim$2. 
This is more clearly illustrated in Fig.~\ref{fig:flux_dist} (left panel), which shows the distribution of neutrino fluxes for sources (real sources, not the simulated sources) passing different TS cuts. As shown in the figure, the reconstructed neutrino fluxes are indeed concentrated within a relatively narrow range (a factor of $\sim 5$, slightly larger than the simple estimate of $2$--$3$ due to moderate variations in the background level across the Northern sky).

In contrast, the luminosity distance of our sample spans 3--4 orders of magnitude (from nearby Seyferts at $z \sim 0.001$ to distant blazars at $z > 1$), corresponding to a factor of $\sim$10$^6$--10$^8$ in $D_L^2$ (see the right panel of Fig.~\ref{fig:flux_dist}). Since luminosity is related to flux by $L = 4\pi D_L^2 F$, the luminosity is dominated by the distance term rather than by flux variations.
The simulated $r$ distributions in Fig.~\ref{fig:sim_r_P_L} further confirm this phenomenon. As the TS threshold increases, the fraction of strong linear correlations produced by pure background fluctuations rises significantly, indicating that distance effects progressively dominate the apparent correlation as the neutrino flux is confined to a narrow range.

This creates an artificial correlation: sources with similar neutrino fluxes (due to the TS cut) but different distances will have neutrino luminosities that scale as $D_L^2$. The X-ray luminosities of these sources also roughly scale as $D_L^2$ (because the X-ray flux is similarly constrained to a narrow range by the narrow energy band of the hard X-ray survey and by the AGN X-ray luminosity function). Since both luminosities share the same $D_L^2$ dependence while the intrinsic flux variations are suppressed by the selection cuts, any sample selected by a flux threshold will exhibit an apparent $L_\nu$--$L_{\rm X}$ correlation even in the absence of an underlying physical connection.

\subsection{The correlation between X-ray and neutrino fluxes}

In the analysis of \citet{Kun2024}, they reported linear correlations in both luminosity and flux spaces for their AGN neutrino emitter candidates. Here we also examine the $F_\nu$--$F_X$ relation for our sample. Fig.~\ref{fig:obs_flux_correlation} shows the $F_\nu$--$F_X$ plane under different TS cuts. The Pearson correlation coefficients in logarithmic flux space are $r_F = 0.07$, $0.10$, $0.20$, and $-0.38$ for TS $>0$, $>2$, $>6$, and $>8$, respectively. The distribution of the Pearson correlation coefficients for 100 random-position simulations are shown in Fig.~\ref{fig:sim_r_P_F}. We can see that neither the real sample nor the Monte Carlo simulations exhibit a significant flux correlation. This indicates that, although the TS cut selection can produce a spurious luminosity correlation, it does not guarantee that the neutrino and X-ray fluxes are correlated.

It is important to contrast this null result with the findings of \citet{Kun2024}. For their pre-selected sample of promising AGN neutrino emitters (six sources), \citet{Kun2024} reported a Pearson coefficient of $0.737$ in flux space. This value suggests a strong flux correlation and lies well above the bulk of the $r_F$ distributions for our simulated samples (Fig.~\ref{fig:sim_r_P_F}). The coexistence of both luminosity and flux correlations in the \citet{Kun2024} sources, whereas the TS/flux cut selection can only produce a spurious luminosity correlation rather than a flux correlation, likely reflects an intrinsic physical connection between the X-ray and neutrino emission for the six sources.
Because the flux correlation eliminates the distance dependence, it indicates that the X-ray and neutrino emission levels from these sources are comparable, both intrinsically and as observed.
Consequently, our analysis does not falsify the physical connection proposed by \citet{Kun2024}. Our results merely show that a luminosity-luminosity correlation, by itself, is insufficient to establish such a connection for a flux-threshold-limited population. To firmly establish such a physical connection, additional evidence is required (for example, a robust flux correlation).
However, it should also be noted that in our simulations, for cases with larger cut thresholds (such as TS $>8$, see the lower-right panel of Fig.~\ref{fig:sim_r_P_F}), there is still a non-negligible probability ($\sim10\%$) of obtaining $r_F>0.737$.
Therefore, the possibility that the flux correlation for the six sources arises from fluctuation cannot be completely ruled out.

\section{Discussion and Conclusion}
\label{sec:discussion}

In this work, we demonstrate that the apparent linear correlation between X-ray and neutrino luminosities in AGN, reported by recent studies, can be fully explained by selection effects rather than physical relationships. Our analysis of 1114 AGN from the \textit{Swift} BAT catalog using 10 years of IceCube data reproduces the correlation for sources with TS$>2$. However, simulations using random sky positions show the same correlation, demonstrating that it is an artifact of the analysis method.
The key issue is that significance-based source selection restricts the neutrino flux to a narrow range, while the luminosity distance spans many orders of magnitude. This causes the luminosity to be dominated by distance rather than intrinsic flux, creating an artificial correlation. Our much larger sample (1114 sources) allows us to clearly see the effect of TS cuts and to perform meaningful simulations.

Our results have important implications for the identification of neutrino sources and the interpretation of multi-messenger correlations. The apparent $L_\nu$--$L_X$ correlation has been used to support a physical connection between neutrino and X-ray production in AGN, and to predict which X-ray sources might be detectable as neutrino emitters. However, our analysis shows that for a flux-limited ensemble, such correlations can arise purely from selection effects and should not be interpreted as evidence for a population-wide physical relationship.
Our results indicate that current neutrino observations are insufficient to establish a quantitative physical relationship between neutrino and X-ray luminosities in AGN for flux-limited ensembles.

This does not mean that there is no physical connection between neutrino and X-ray emission in AGN. Theoretical models do predict correlations under certain assumptions about the emission mechanisms \cite{Murase2016,Murase2020}. 
It is also noteworthy that the six sources studied by \citet{Kun2024} satisfy not only the luminosity--luminosity correlation but also exhibit a flux--flux correlation. Our simulations demonstrate that while selection effects can readily produce a spurious $L_\nu$--$L_X$ correlation, they cannot generate a fake $F_\nu$--$F_X$ correlation. Therefore, the X-ray and neutrino emission from the six sources in \citet{Kun2024} may indeed be physically linked. 
Our results only caution that the establishment of a physical connection must be supported by evidence beyond the luminosity--luminosity relation alone. 
Undoubtedly establishing such correlations may require more sensitive neutrino observatories (such as IceCube-Gen2 \cite{IceCubeGen2}, TRIDENT \cite{TRIDENT}, HUNT \cite{HUNT}) and careful treatment of selection effects and detection thresholds.

\begin{acknowledgments}
We thank the IceCube Collaboration for making their 10-yr muon-track data publicly available. This work used the \texttt{skyllh} software package for neutrino point-source analysis.
This work is supported by the Innovation Project of Guangxi Graduate Education (YCBZ2024060).
\end{acknowledgments}

\bibliographystyle{apsrev4-1-lyf}
\bibliography{manuscript}{}

\end{document}